# Inferences about the distribution, merger rate, and evolutionary processes of compact binaries from gravitational wave observations

M.S. *Master of Science*

in Astrophysical Sciences and Technology

*Daniel Wysocki*

School of Physics and Astronomy

Rochester Institute of Technology

Rochester, New York

September 2017

ASTROPHYSICAL SCIENCES AND TECHNOLOGY
COLLEGE OF SCIENCE
ROCHESTER INSTITUTE OF TECHNOLOGY
ROCHESTER, NEW YORK

M.S. THESIS DEFENSE

---

Candidate: ....................................................

Thesis Title: ..............................................................

Adviser: ......................................................

Date of defense: ................................

The candidate's M.S. Thesis has been reviewed by the undersigned. The Thesis

(a) is acceptable, as presented.

(b) is acceptable, subject to minor amendments.

(c) is not acceptable in its current form.

*Written details of required amendments or improvements have been provided to the candidate.*

Committee:

Dr. Jason Nordhaus, Committee Member

Dr. John Whelan, Committee Member

Dr. Richard O'Shaughnessy, Thesis Advisor

Please submit form to AST Graduate Program Coordinator

ASTROPHYSICAL SCIENCES AND TECHNOLOGY
COLLEGE OF SCIENCE
ROCHESTER INSTITUTE OF TECHNOLOGY
ROCHESTER, NEW YORK

# CERTIFICATE OF APPROVAL

---

**M.S. DEGREE THESIS**

The M.S. Degree Thesis of *Daniel Wysocki* has been examined and approved by the thesis committee as satisfactory for the thesis requirement for the M.S. degree in Astrophysical Sciences and Technology.

________________________________________
Dr. Jason Nordhaus, Committee Member

________________________________________
Dr. John Whelan, Committee Member

________________________________________
Dr. Richard O'Shaughnessy, Thesis Advisor

Date __________________________________

INFERENCES ABOUT THE DISTRIBUTION, MERGER RATE, AND EVOLUTIONARY PROCESSES OF COMPACT BINARIES FROM GRAVITATIONAL WAVE OBSERVATIONS

By

*Daniel Wysocki*

A dissertation submitted in partial fulfillment of the requirements for the degree of M.S. in Astrophysical Sciences and Technology, in the College of Science, Rochester Institute of Technology.

September, 2017

Approved by

______________________________ ______________

Dr. Joel Kastner Date

Director, Astrophysical Sciences and Technology

# Declaration

I, DANIEL M. WYSOCKI ("the Author"), declare that no part of this thesis is substantially the same as any that has been submitted for a degree or diploma at the Rochester Institute of Technology or any other University. I further declare that the work in Chapters 3 and 4 are entirely my own; Chapters 1, 2, and 5 draw in parts on work done with collaborators, including contributions from other authors. Those who have contributed scientific or other collaborative insights are fully credited in this thesis, and all prior work upon which this thesis builds is cited appropriately throughout the text. ***NOT YET:*** *This thesis was successfully defended in Rochester, NY, USA on the 27th of September, 2017.*

Modified portions of this thesis will be published by the author and his advisor, in a journal(s) yet to be determined.

- **Chapter 2** is based on the paper [1], entitled *Explaining LIGO's observations via isolated binary evolution with natal kicks* authored by D. Wysocki, D. Gerosa, R. O'Shaughnessy, K. Belczynski, W. Gladysz, E. Berti, M. Kesden, and D. Holz., arxiv:1709.01943

## Abstract

We are living through the dawn of the era of gravitational wave astronomy. Our first glances through this new window upon the sky has revealed a new population of objects. Since it first began observing in late 2015, the advanced Laser Interferometer Gravitational-Wave Observatory (LIGO) has detected gravitational waves three times, along with an additional strong candidate [2, 3] – and there shall be orders of magnitude more in the years to come. In all four cases, the waveform's signature is consistent with general relativity's predictions for the merging of two black holes. Through parameter estimation studies, estimates on features such as the black holes' masses and spins have been determined. At least two of the black hole pairs lie above the mass range spanned by comparable black holes observed through traditional means [4, 3]. This suggests they constitute a separate population, either too elusive or rare to be found with traditional telescopes.

The most natural questions to ask about these black holes – how did they form, how many of them are there, and how can they be categorized – remain open ended. We know black holes can form when massive stars die, so it's most natural to claim stars as their progenitors. Since we now know black holes merge into larger black holes, could it be the case that they formed from previous mergers? [5, 6] Were the two black holes part of a binary from their birth [1], or did they become coupled later on in life [7]? The measurements provided by LIGO can help answer these questions and more.

Throughout this thesis, I will describe and demonstrate results from a number of novel methods whose purpose is to better understand these black holes and their progenitors. At their heart, these methods give answers to a few, critical questions. a) What is the overall rate at which these objects merge? b) What is the range of values these objects' properties can take, and how are they distributed? c) Given a number of physical models, how can we evaluate the performance of each relative to the others, and determine which gives the best description of reality? We accomplish this through a host of statistical methods.

# Contents

# List of Figures

# List of Tables

# Chapter 1

# Introduction

The dawn of gravitational wave (GW) astronomy took place on September 14, 2015, when the Laser Interferometer Gravitational-Wave Observatory (LIGO) detected the gravitational radiation from two black holes merging [4]. This event has been named GW150914, after its discovery date. Since then, LIGO has detected two other binary black hole mergers, GW151226 [8] and GW170104 [3], as well as one other binary black hole candidate at only the $\sim 90\%$ confidence level, LVT151012 [9] [1]. As LIGO continues to observe and undergo sensitivity upgrades over the next few years, the current $\sim 3.9$ detections will grow into the hundreds and even thousands.

It is clear from the current merger rate estimates [10, 9, 3] that the Universe is filled with these binary black holes. While a number of low mass black hole binaries were found with electromagnetic (EM) observations prior to LIGO [11], the higher mass events, GW150914 [4] and GW170104 [3], clearly belong to a different sub-population, indicating that those surveys are incomplete. Furthermore, GW observations allow us to measure parameters unobtainable from EM observations, such as spins [12]. Also, considering that black holes do not emit light (hence the name *black* holes), GW's are simply a more naturally suited tool to observe them, as they are emitted directly from the black holes, as opposed to being a result of external matter interacting with them. With all of this in mind, this population of objects is worthy

[1] LVT stands for LIGO-Virgo trigger. Whereas confident detections are labeled GW for gravitational wave, LVT's are highly interesting candidates, below the $5\sigma$ level required to claim a detection.

of study, and is most effectively done through GW observations.

Natural questions arise, such as:

1. Where do these black holes come from, and how did they evolve?

2. What are the overall properties of their populations, including the distribution of their masses and spins?

3. How many of these systems exist in the Universe, and how frequently do they merge?

All of these and more can be answered, at least to some extent, with enough observed GW events, the right statistical methodology, and developments in astrophysical simulations. In addition, answering or constraining Questions 2 and 3 can help answer 1, as any proposed solution to 1 will make a prediction about 2 and 3 that can be validated or falsified.

There two main paradigms of solutions to Question 1. They state the the binary black holes observed by LIGO are either:

1. the remnants of binary star systems, where the two stars were produced from the same circumbinary disk, and eventually the two stars collapsed into black holes [13, 1]. This is commonly referred to as isolated formation.

2. produced by two objects that were born separately (either black holes or stars that evolve into black holes later on) and eventually became coupled through chance interactions in a dense cluster of objects, such as globular clusters. [7]. This is commonly referred to as dynamical formation.

Also falling under Solution 2 is the possibility that the black holes are themselves the byproducts of previous black hole mergers [5, 6].

The reality is likely a mixture of both paradigms, since we know stars form binaries and that massive stars form black holes, and the Universe is big enough that some black holes must form binaries through chance interactions. The real question is to what extent does each formation channel contribute to the population? Does one dominate the other, or do they contribute comparably? This is the current hot topic in GW astrophysics.

Solution 1 is of particular interest, because it means we can test the predictions of specific binary stellar evolution models. There are still many fundamental questions about the evolution of binary stars which remain unanswered, and thus Chapter 2 of this thesis is devoted to comparing a set of binary evolution models to LIGO's 3.9 detections. This chapter was produced from a paper we are in the process of publishing [1]. We also recently published a less detailed paper on this same subject, focusing on explaining the GW151226 event, though my contribution was not significant enough to warrant inclusion in this thesis [13].

In Chapter 3, we describe a method for answering Questions 2 and 3 simultaneously, using Bayesian inference and parametric models. Then in Chapter 4, we describe a method specifically for answering Question 2 using weakly parametric mixtures of Gaussian distributions.

The chapter progression starts with the strongest assumptions (specific stellar evolution models assumed as the progenitors in Chapter 2) and progressively weakens the assumptions, moving from strongly (Chapter 3) to weakly (Chapter 4) parametric statistical descriptions of the populations.

# Chapter 2

# Method I: Mixture of physical models

In this chapter, we infer the parameter distribution of binary black holes given a discrete grid of physical models. On this grid we vary the natal spin magnitudes of the two black holes, the strength of supernova kicks ($\sigma_{\rm kick}$), and the tidal physics model. The entirety of this chapter is taken a paper which has been submitted to Phys. Rev. Dand is currently awaiting publication [1].

## 2.1 Introduction

The discovery and interpretation of gravitational waves (GW) from coalescing binaries [4] has initiated a revolution in astronomy [14]. Several hundred more detections are expected over the next five years [10, 9, 3]. Already, the properties of the sources responsible – the inferred event rates, masses, and spins – have confronted other observations of black hole (BH) masses and spins [9], challenged previous formation scenarios [14, 9], and inspired new models [15, 16, 17, 18] and insights [19, 20] into the evolution of massive stars and the observationally accessible gravitational waves they emit [21, 22]. Over the next several years, our understanding of the lives and deaths of massive stars over cosmic time will be transformed by the identification and interpretation of the population(s) responsible for coalescing binaries, with and without counterparts, because measurements will enable robust tests to distinguish between formation scenarios with present [23, 13] and future instruments [24, 25], both coarsely and with high precision. In this work, we demonstrate the power of gravitational wave measurements to constrain how BHs form, within the context of one formation scenario for binary BHs: the isolated evolution of pairs of stars [26, 27, 28, 29, 30, 31, 32, 33, 34, 35, 36].

Within the context of that model, we focus our attention on the one feature whose unique impacts might be most observationally accessible: BH natal kicks. Observations strongly suggest that when compact objects like neutron stars are formed after the death of a massive star, their birth can impart significant linear momentum or "kick". For example, observations of pulsars in our galaxy suggest birth velocity changes as high as $v_k \sim 450$ km/s [37]. These impulsive momentum changes impact the binary's intrinsic orbit and stability, changing the orbital parameters like semimajor axis and orbital plane [38, 39], as well as causing the center of mass of the remnant BH binary (if still bound) to recoil at a smaller but still appreciable velocity. While no single compelling and unambiguous observation can be explained only with a BH natal kick, the assumption of small but nonzero BH natal kicks provides a natural explanation for several observations, including the posterior spin-orbit misalignment distribution of GW151226 and the galactic X-ray binary misalignment [40, 41, 42, 43] and recoil velocity [44, 45, 46, 47, 48, 49]. Modest BH natal kicks can be produced by, for example, suitable neutrino-driven supernova engines; see, e.g., [50] and references therein.

We compare binary formation models with different BH natal kick prescriptions to LIGO observations of binary black holes. Along with [50], our calculation is one of the first to perform this comparison while changing a single, physically well-defined and astrophysically interesting parameter: the BH natal kick strength. It is the first to self-consistently draw inferences about binary evolution physics by comparing observations simultaneously to the predicted detection rate; binary BH masses; and binary BH spins, accounting for both magnitude and misalignment.

This comparison is important because BH natal kicks introduce two complementary and unusually distinctive effects on the binary BHs that LIGO detects. On the one hand, strong BH natal kicks will frequently disrupt possible progenitor binary systems. As the strength of BH natal kicks increases, the expected number of coalescing binary BHs drops precipitously [27, 51, 28]. On the basis of observations to date, BH natal kicks drawn from a distribution with one-dimensional velocity dispersion $\sigma$ greater than 265 km/s are disfavored [34]. On the other hand, BH natal kicks will tilt the orbital plane, misaligning the orbital angular momentum from the black hole's natal spin direction – assumed parallel to the progenitor binary's orbital angular momentum [39, 13]. The imprint of these natal kicks on the binary's dynamics is preserved over the aeons between the BH-BH binary's formation and its final coalescence [52, 38, 53, 54]. The outgoing radiation from each merger contains information about the coalescing binary's spin (see, e.g., [12, 55, 56] and references therein), including conserved constants that directly reflect the progenitor binary's state [57, 58]. Several studies have demonstrated that the imprint of processes that misalign BH spins and the orbit can be disentangled [59, 7, 60].

In this work, we show that LIGO's observations of binary black holes can be easily explained in the context of isolated binary evolution, if BH natal kicks act with the (modest) strength to misalign the orbital plane from the initial spin directions (presumed aligned). In this approach, the absence of large aligned spins either reflects fortuitous but nonrepresentative observations or low natal BH spins. A companion study by **(author?)** [50] describes an alternative, equally plausible explanation: the BH natal spin depends on the progenitor, such that the most massive BHs are born with low natal spins. A longer companion study by **(author?)** [61] will describe the properties and precessing dynamics of this population in greater detail.

This paper is organized as follows. First, in Section 2.2 we describe the entire process used to generate and characterize a detection-weighted populations of precessing binary BHs, evaluated using different assumptions about BH natal kicks. As described in Section 2.2.1, we adopt previously studied binary evolution calculations to determine how frequently compact binaries merge throughout the universe. In Section 2.2.2, we describe how we evolve the binary's precessing BH spins starting from just after it forms until it enters the LIGO band. In Section 2.2.3, we describe the parameters we use to characterize each binary: the component masses and spins, evaluated after evolving the BH binary according to the process described in Section 2.2.2. To enable direct comparison with observations, we convert from detection-weighted samples – the output of our binary evolution model – to a smoothed approximation, allowing us to draw inferences about the relative likelihood of generic binary parameters. In Section 2.3 we compare these smoothed models for compact binary formation against LIGO's observations to date. We summarize our conclusions in Section 2.6. In Appendix B.1 we describe the technique we use to approximate each of our binary evolution simulations. In Appendix B.2, we provide technical details of the underlying statistical techniques we use to compare these approximations to LIGO observations. To facilitate exploration of alternative assumptions about natal spins and kicks, we have made publicly available all of the marginalized likelihoods evaluated in this work, as supplementary material.

## 2.2 Estimating the observed population of coalescing binary black holes

### 2.2.1 Forming compact binaries over cosmic time

Our binary evolution calculations are performed with the `StarTrack` isolated binary evolution code [28, 63], with updated calculation of common-envelope physics [31], compact remnant masses [64], and pair instability supernovae [62]. Using this code, we generate a synthetic universe of (weighted) binaries by Monte Carlo [65]. Our calculations account for the time- and metallicity- dependent star formation history of the universe, by using a grid of **32** different choices for stellar metallicity. As shown in Table 2.1, we create synthetic universes using the same assumptions (M10) adopted by default in previous studies [62, 34, 66]. Again as in previous work, we explore a one-parameter family of simulations that adopt different assumptions about BH natal kicks (M13-M18). Each new model assumes all BHs receive natal kicks drawn from the same Maxwellian distribution, with one-dimensional velocity distribution parameterized by $\sigma$ (a quantity which changes from model to model). In the M10 model used for reference, BH kicks are also drawn from a Maxwellian distribution, but suppressed by the fraction of ejected material that is retained (i.e., does not escape to infinity, instead being accreted by the BH). Because the progenitors of the most massive BHs do not, in our calcu-lations, eject significant mass to infinity, the heaviest BHs formed in this "fallback-suppressed kick" scenario receive nearly or exactly zero natal kicks.

| Name | $\sigma$ (km/s) | $D_{KL}(M)$ | $D_{KL}(m_1, m_2)$ |
|---|---|---|---|
| M10 | | 0.02 | 0.21 |
| M18 | 25 | 0.006 | 0.094 |
| M17 | 50 | 0 | 0 |
| M16 | 70 | 0.016 | 0.28 |
| M15 | 130 | 0.1 | 1.26 |
| M14 | 200 | 0.17 | 1.56 |
| M13 | 265 | 0.40 | 2.1 |

Table 2.1: Properties of the formation scenarios adopted in this work. The first column indicates the model calculation name, using the convention of other work [34, 62]. The second column provides the kick distribution width. Model M10 adopts mass-dependent, fallback suppressed BH natal kicks. For the BH population examined here, these natal kicks are effectively zero for massive BHs; see, e.g., [23]. The remaining scenarios adopt a mass-independent Maxwellian natal kick distribution characterized by the 1-d velocity dispersion $\sigma$, as described in the text. The third column quantifies how much the mass distribution changes as we change $\sigma$. To be concrete, we compare the (source frame) total mass distributions for the BH-BH binaries LIGO is expected to detect, using a KL divergence [Eq. (2.4)]. If $p(M|\alpha)$ denotes the mass distribution for $\alpha =$ M10, M18, M17, $\ldots$, and $\alpha_*$ denotes M17, then the third column is the KL divergence $D_{KL}(M, \alpha) = \int dM p(M|\alpha) \ln[p(M|\alpha)/p(M|\alpha_*)]$. The fourth column is the KL divergence using the joint distribution of both binary masses: $D_{KL}(m_1, m_2|\alpha) = \int dm_1 dm_2 p(m_1, m_2|\alpha) \ln[p(m_1, m_2|\alpha)/p(m_1, m_2|\alpha_*)]$. Because M10 adopts fallback-suppressed natal kicks, while the remaining models assume fallback-independent natal kicks, we use the special symbol  to refer to M10 in subsequent plots and figures.

These synthetic universes consist of weighted BH-BH mergers (indexed by $i$), each one acting a proxy for a part of the overall merger rate density in its local volume [67, 68]. As our synthetic universe resamples from the same set of **32** choices for stellar metallicity, the same evolutionary trajectory appears many times, each at different redshifts and reflecting the relative probability of star formation at different times.

The underlying binary evolution calculations performed by `StarTrack` effectively do not depend on BH spins at any stage.[1] We therefore have the freedom to re-use each calculation above with any BH natal spin prescription whatsoever. Unlike **(author?)** [66], we do not adopt a physically-motivated and mass-dependent BH natal spin, to allow us to explore all of the possibilities that nature might allow. Instead, we treat the birth spin for each BH as a parameter, assigning spins $\chi_1$ and $\chi_2$ to each black hole at birth. For simplicity and without loss of generality, for each event we assume a fixed BH spin for the first-born ($\chi_1 = |\mathbf{S}_1|/m_1^2$) and a potentially different spin for the second-born ($\chi_2 = |\mathbf{S}_2|/m_2^2$) BH. Both choices of fixed spin are parameters. By carrying out our calculations on a discrete grid in $\chi_1, \chi_2$ for each event – here, we use $\chi_{1,2} = 0.1 \ldots 1$ – we encompass a wide range of possible choices for progenitor spins, allowing us to explore arbitrary (discrete) natal spin distributions. For comparison, [7] adopted a fixed natal spin $\chi_i = 0.7$ for all BHs. Our choices for BH natal spin distributions are restricted only by our choice of discrete spins. Our model is also implicitly limited by requiring all BHs have natal spins drawn from the same mass-independent distributions. By design, our calculation did not include enough degrees of freedom to enable the natal spin distribution to change with mass, as was done for example in [50].

[1] The response of the BH's mass and spin to accretion depends on the BH's spin. We adopt a standard procedure whereby the BH accretes from the innermost stable circular orbit. In our binary evolution code, this spin evolution is implemented directly via an ODE based on (prograde, aligned) ISCO accretion as in [69], though the general solution is provided in [70] and applied since, e.g., in [71, 72]. For the purposes of calculating the final BH mass from the natal mass and its accretion history, we adopted a BH natal spin of $\chi = 0.5$; however, relatively little mass is accreted and the choice of spin has a highly subdominant effect on the BH's evolution.

We assume the progenitor stellar binary is comprised of stars whose spin axes are aligned with the orbital angular momentum, reflecting natal or tidal [73, 74] alignment (but cf. [75]). After the first supernova, several processes could realign the stellar or BH spin with the orbital plane, including mass accretion onto the BH and tidal dissipation in the star. Following **(author?)** [38], we consider two possibilities. In our default scenario ("no tides"), spin-orbit alignment is only influenced by BH natal kicks. In the other scenario ("tides"), tidal dissipation will cause the stellar spin in stellar-BH binaries to align parallel to the orbital plane. In the "tides" scenario, the second-born stellar spin is aligned with the orbital angular momentum prior to the second SN. Following [38], the "tides" scenario assumes alignment always occurs for merging BH-BH binaries, independent of the specific evolutionary trajectory involved (e.g., binary separation); cf. the discussion in [66]. In both formation scenarios, we do not allow mass accretion onto the BH to change the BH's spin direction. Given the extremely small amount of mass accreted during either conventional or common-envelope mass transfer, even disk warps and the Bardeen-Petterson effect should not allow the BH spin direction to evolve [76, 77, 78, 79]. For coalescing BH-BH binaries the second SN often occurs when the binary is in a tight orbit, with high orbital speed, and thus less effect on spin-orbit misalignment [39, 13]. Therefore, in the "tides" scenario, the second-born BH's spin is more frequently nearly aligned with the final orbital plane, even for large BH natal kicks.

| Formation mechanism | Fraction |
|---|---|
| MT1(2-1) MT1(4-1) SN1 CE2(14-4;14-7) SN2 | 0.261 |
| MT1(4-4) CE2(7-4;7-7) SN1 SN2 | 0.234 |
| MT1(4-1) SN1 CE2(14-4;14-7) SN2 | 0.140 |
| MT1(2-1) SN1 CE2(14-4;14-7) SN2 | 0.075 |
| MT1(4-4) CE2(4-4;7-7) SN1 SN2 | 0.071 |
| MT1(2-1) SN1 MT2(14-2) SN2 | 0.037 |
| CE1(4-1;7-1) SN1 MT2(14-2) SN2 | 0.028 |
| CE1(4-1;7-1) SN1 CE2(14-4;14-7) SN2 | 0.020 |
| CE1(4-1;7-1) CE2(7-4;7-7) SN1 SN2 | 0.014 |
| MT1(4-4) CE12(4-4;7-7) SN2 SN1 | 0.014 |
| SN1 CE2(14-4;14-7) SN2 | 0.014 |
| Other channels | 0.16 |

Table 2.2: The most significant formation scenarios and fraction of detected binaries formed from that channel, for the M15 model. While many of the coalescing BH-BH binaries form via a BH-star binary undergoing some form of stellar mass transfer or interaction, a significant fraction of binaries form without any Roche lobe overflow mass transfer after the first SN. In this example, in the second channel alone more than **23%** of binaries form without interaction after the first SN. (The remaining formation channels account for **16%** of the probability.) In this notation, integers in braces characterize the types of the stellar system in the binary; the prefix refers to different phases of stellar interaction (e.g., MT denotes "mass transfer," SN denotes "supernova," and CE denotes "common envelope evolution"); and the last integer SN$x$ indicates whether the initial primary star (1) or initial secondary star (2) has collapsed and/or exploded to form a BH. [Some of our BHs are formed without luminous explosions; we use SN to denote the death of a massive star and the formation of a compact object.] A detailed description of these formation channels and stellar types notation is provided in [28, 63]; in this shorthand, 1 denotes a main sequence star; 2 denotes a Hertzprung gap star; 4 denotes a core heium burning star; 7 denotes a main sequence naked helium star; and 14 denotes a black hole.

### 2.2.2 Evolving from birth until merger

The procedure above produces a synthetic universe of binary BHs, providing binary masses, spins, and orbits just after the second BH is born. Millions to billions of years must pass before these binaries coalesce, during which time the orbital and BH spin angular momenta precess substantially [53, 54]. We use precession-averaged 2PN precessional dynamics, as implemented in `precession` [80], to evolve the spins from birth until the binary BH orbital frequency is 10Hz (i.e., until the GW frequency is 20Hz); see [61] for details. When identifying initial conditions, we assume the binary has already efficiently circularized. When identifying the final separation, we only use the Newtonian-order relationship between separation and orbital frequency. The `precession` code is publicly available at github.com/dgerosa/precession.

### 2.2.3 Characterizing the observed distribution of binaries

At the fiducial reference frequency adopted in this work (20Hz), a binary BH is characterized by its component masses and its (instantaneous) BH spins $\mathbf{S}_{1,2}$. For the heavy BHs of current interest to LIGO, the principal effect of BH spin on the orbit and emitted radiation occurs through the spin combination

$$\begin{aligned}\chi_{\rm eff} &= (\mathbf{S}_1/m_1 + \mathbf{S}_2/m_2)\cdot\hat{\mathbf{L}}/(m_1+m_2) \\ &= (\chi_1 m_1\cos\theta_1 + \chi_2 m_2\cos\theta_2)/(m_1+m_2), \end{aligned} \tag{2.1}$$

where $\theta_{1,2}$ denote the angles between the orbital angular momentum and the component BH spins. That said, depending on the duration and complexity of the source responsible, GW measurements may also provide additional constraints on the underlying spin directions themselves [57], including on the spin-orbit misalignment angles $\theta_{1,2}$. For the purposes of this work, we will be interested in the (source-frame) binary masses $m_1, m_2$ and the spin parameters $\chi_{\rm eff}, \theta_1, \theta_2$, as an approximate characterization of the most observationally accessible degrees of freedom; cf. **(author?)** [81], which used $\theta_{1,2}$, and **(author?)** [57], which used $\theta_{1,2}$ and the angle $\Delta\Phi$ between the spins' projection onto the orbital plane. In particular, $\Delta\Phi$ is well-known to contain valuable information [38] and be observationally accessible [57]. At present, the preferred model adopted for parameter inference, known as IMRPhenomP, does not incorporate the necessary degree of freedom [82], so we cannot incorporate its effect here. With additional and more informative binary black hole observations, however, our method should be extended to employ all of the spin degrees of freedom, particularly $\Delta\Phi$. As input, this extension will require inference results that incorporate the effect of two two precessing spins, either by using semianalytical models [83, 84, 85] or by using numerical relativity [56].

We adopt a conventional model for LIGO's sensitivity to a population of binary BHs [86, 33, 10]. In this approach, LIGO's sensitivity is limited by the second-most-sensitive interferometer, using a detection threshold signal-to-noise ratio $\rho = 8$ and the fiducial detector sensitivity reported for O1 [9]. This sensitivity model is a good approximation to the performance reported for both in O1 and early in O2 [3]. Following [33, 47], we use the quantity $r_i$ [Eq. (8) in [47]] to account for the contribution of this binary to LIGO's detection rate in our synthetic universe, accounting for the size of the universe at the time the binary coalesces and LIGO's orientation-dependent sensitivity. For simplicity and following previous work [33, 10], we estimate the detection probability without accounting for the effects of BH spin. Previous studies have used this detection-weighted procedure to evaluate and report on the expected distribution of binary BHs detected by LIGO [34, 62, 66]. Since the same binary evolution $A$ occurs many times in our synthetic universe, we simplify our results by computing one overall detection rate $r_A = \sum_{i \in A} r_i$ for each evolution. When this procedure is performed, relatively few distinct binary evolutions $A$ have significant weight. While our synthetic universe contains millions of binaries, only $O(10^4)$ distinct BH-BH binaries are significant in our final results for each of the formation scenarios listed in Table 2.1. Figure 2.1 illustrates the expected detected number versus assumed BH natal kick strength.

The significant BH natal kicks adopted in all of our formation scenarios (except M10) frequently produce significant spin-orbit misaligment. Figure 2.2 shows that strong misalignment occurs ubiquitously, even for small BH natal kicks; see [61] for more details. This strong spin-orbit misalignment distribution produces an array of observationally accessible signatures, most notably via an invariably wide distribution of $\chi_{\text{eff}}$. In [61] the distribution was constructed for all of our models, finding that (except for M10) considerable support exists for $\chi_{\text{eff}} < 0$. Our calculation is fully consistent with the limited initial exploration reported in **(author?)** [23], which claimed $\chi_{\text{eff}} < 0$ was implausible except for extremely large natal kicks. Their collection of calculations explored fallback-suppressed kicks (e.g., equivalent to our model M10); adopted natal kicks larger than we explored here; or adopted mass-dependent natal kicks. We show that significant spin-orbit misalignment is plausible if all BHs – even massive ones – receive a modest natal kick. BH natal kicks therefore provide a robust mechanism to explain the observed $\chi_{\rm eff}$ and spin-orbit misalignments reported by LIGO for its first few detections.

The procedure described above *samples* a synthetic universe and synthetic observations by LIGO. However, to compare to LIGO's observations, we need to be able to assess the likelihood of *generic* binaries according to our formation scenario, extrapolating between what we have simulated. We therefore estimate the merger rate distribution as a function of binary masses, spins, and spin-orbit misalignments. Our estimate uses a carefully calibrated Gaussian mixture model, with special tuning as needed to replicate sharp features in our mass and misalignment distribution; see Appendix B.1 for details.

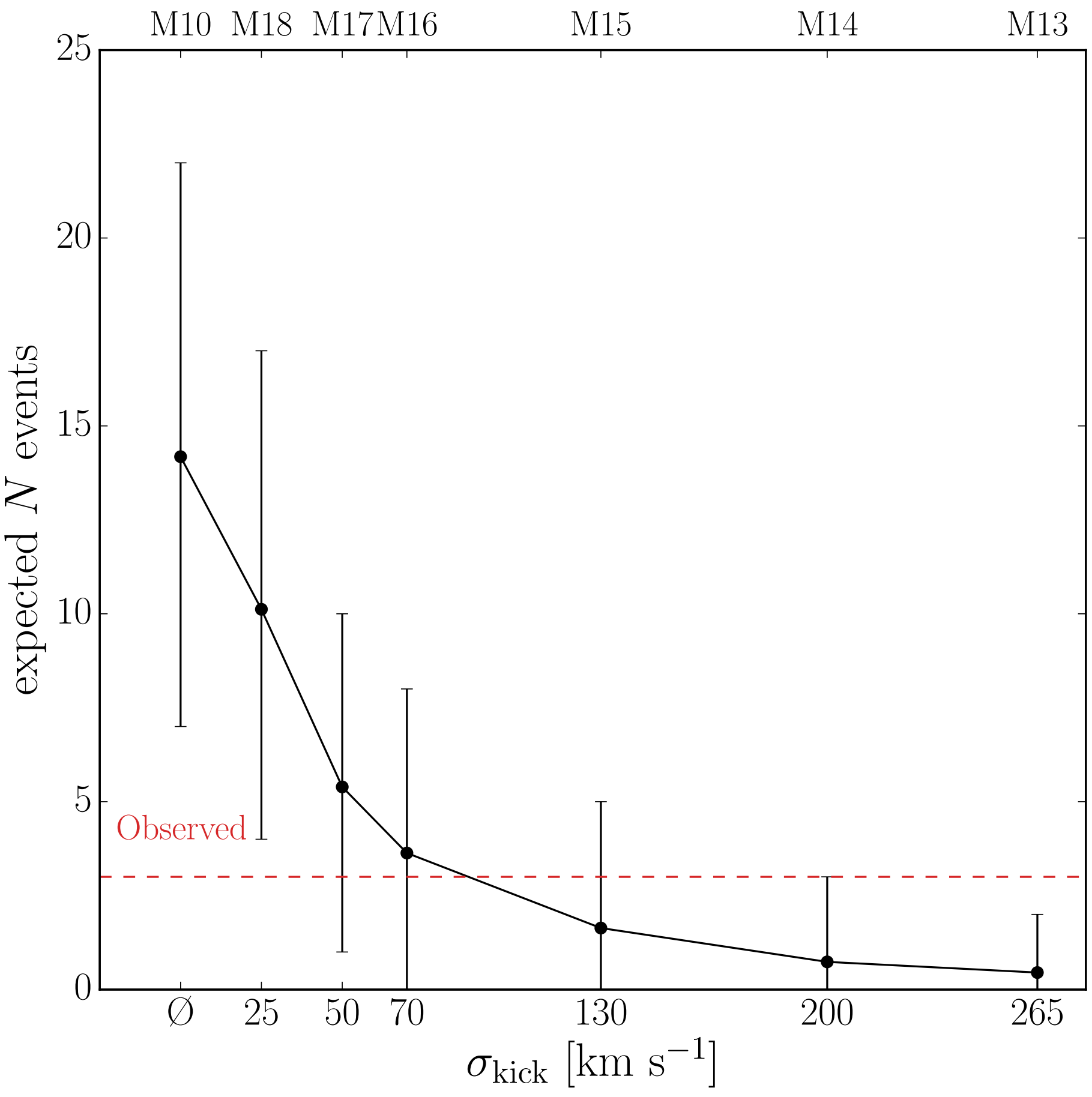

Figure 2.1: **Expected number of events versus kick strength**: Expected number of BH merger detections predicted at LIGO's O1 sensitivity and for the duration of O1 by our formation scenarios. The predicted number of events decreases rapidly as a function of the BH natal kick. Also shown is the 95% confidence interval, assuming Poisson distribution with mean predicted by our model. This purely statistical error bar does not account for any model systematics (e.g., in the overall star formation rate and metallicity history of the universe). The horizontal red dashed line corresponds to the number (3) of observations reported in O1 [9].

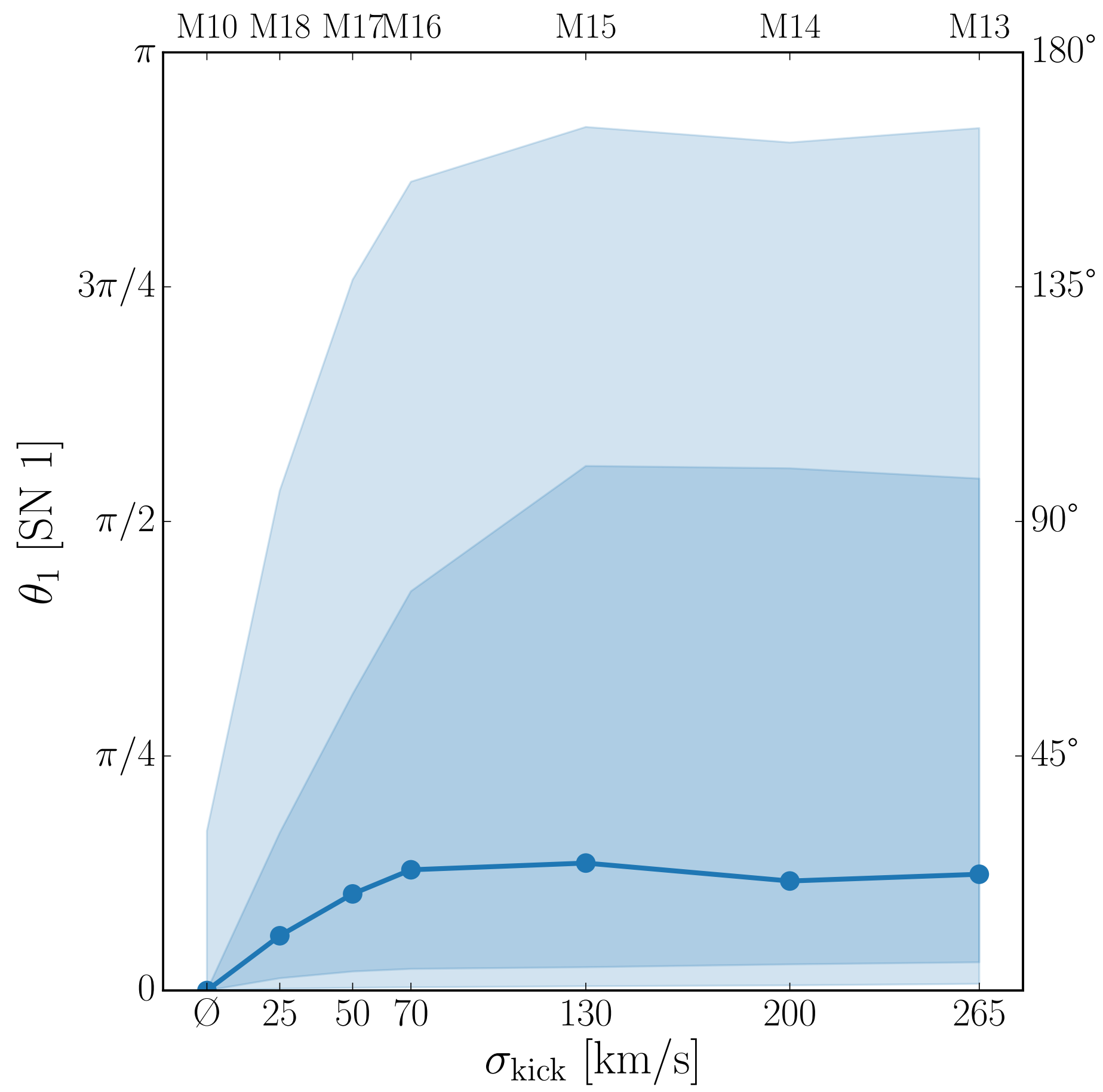

Figure 2.2: **Spin-orbit misalignment versus kick strength**: The misalignment $\theta_{1,SN1}$ after the first SN event, as a function of the characteristic BH natal kick $\sigma$. (Note $\theta_{1,SN1}$ should be distinguished from $\theta_1$ described in the text: $\theta_1$ is the angle between the more massive BH and the orbital angular momentum, at 20 Hz.) The solid line shows the median value; shaded region shows the 68% and 95% confidence intervals.

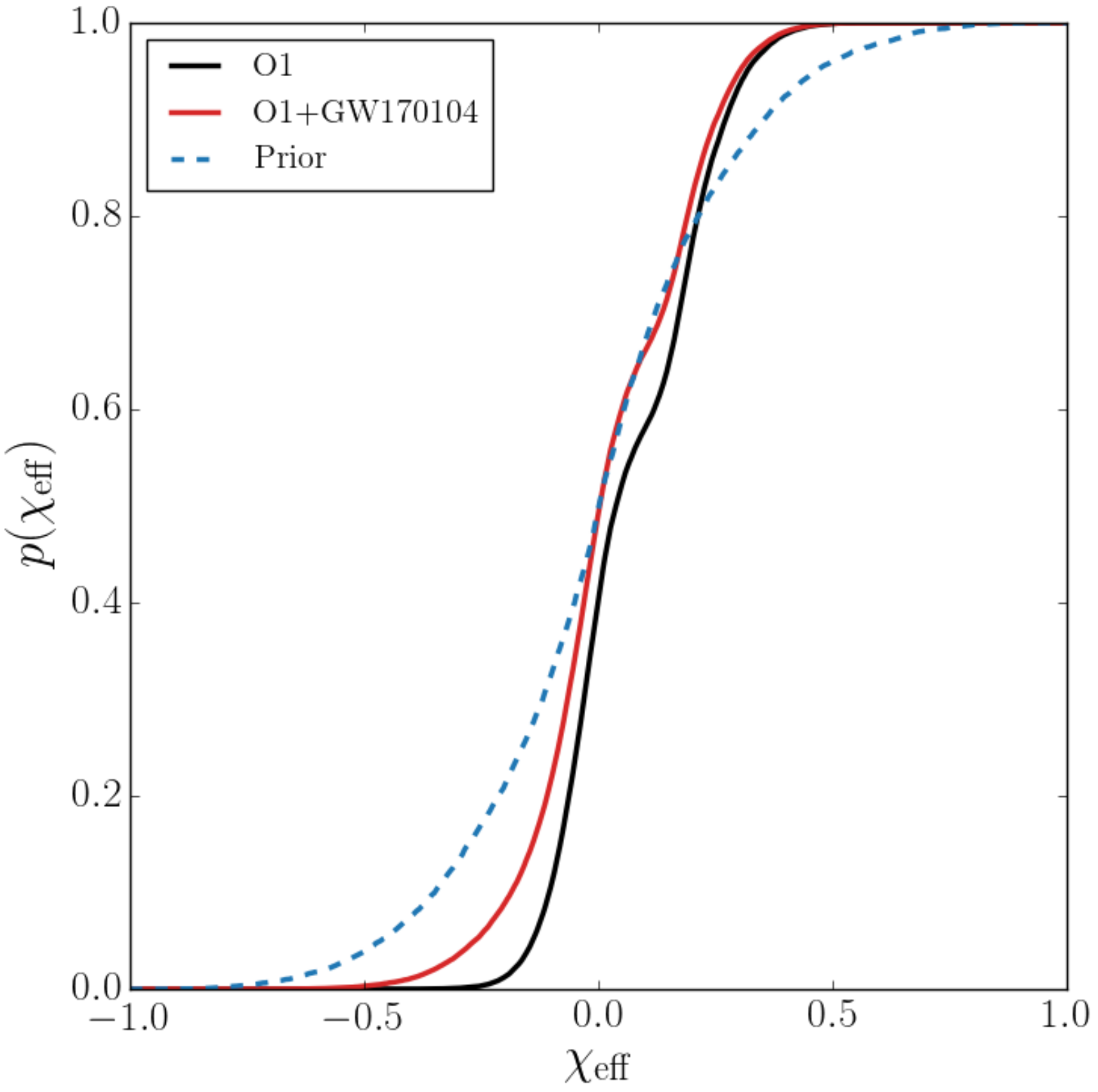

Figure 2.3: **Empirical cumulative distribution function for $\chi_{\rm eff}$:** The solid blue line shows the conventional prior distribution for $\chi_{\rm eff}$, generated by selecting masses uniformly in $m_{1,2} \geq 1M_\odot$, $m_{1,2} \leq 100M_\odot$, $m_1+m_2 < 100M_\odot$, and isotropic spins generated independently and uniformly in magnitude. This prior was adopted when analyzing all LIGO events. The solid black line shows the empirical cumulative distribution for $\chi_{\rm eff}$, derived from the collection of events $\alpha$ = GW150914, GW151226, and LVT151012 via their posterior cumulative distributions $P_\alpha(\chi_{\rm eff})$ via $P(\chi_{\rm eff}) = \sum_\alpha P_\alpha(\chi_{\rm eff})/3$. In this curve, the posterior distributions are provided by LIGO's full O1 analysis results [9], as described in the text. The solid red line shows the corresponding result when GW170104 is included. The approximate posterior distribution for GW170104 is based on published results, as described in Appendix B.3.

## 2.3 Comparison with gravitational wave observations

### 2.3.1 Gravitational wave observations of binary black holes

During its first observing run of $T_1 = 48.6$ days, LIGO has reported the observation of three BH-BH mergers: GW150914, LVT151012, and GW151226 [4, 8, 9]. In an analysis of $T_2 = 11$ days of data from its second observing run, at comparable sensitivity, LIGO has since reported the observation of another binary BH: GW170104 [3]. To draw out more insight from each observation, rather than use the coarse summary statistics LIGO provides in tabular form, we employ the underlying posterior parameter distribution estimates provided by the LIGO Scientific Collaboration for the three O1 events [9, 55, 12]. For GW170104, we instead adopt an approximate posterior distribution described in Appendix B.3 based solely on reported tabular results; that said, we are confident that this approximation makes no difference to our conclusions. For each event, for brevity indexed by an integer $n = 1, 2, 3, \ldots, N$, these estimates are generated by comparing a proposed gravitational wave source $x$ with the corresponding stretch of gravitational wave data $d$ using a (Gaussian) likelihood function $p(d|x)$ that accounts for the frequency-dependent sensitivity of the detector (see, e.g., [12, 55, 56] and references therein). In this expression $x$ is shorthand for the 15 parameters needed to fully specify a quasicircular BH-BH binary in space and time, relative to our instrument; and $d$ denotes all the gravitational wave data from all of LIGO's instruments. This analysis adopts prior assumptions about the relative likelihood of different progenitor binaries $p_{\rm ref}(x)$: equally likely to have any pair of component masses, any spin direction, any spin magnitude, any orientation, and any point in spacetime (i.e., uniform in comoving volume). Then, using standard Bayesian tools [55, 12], the LIGO analysis produced a sequence of independent, identically distributed samples $x_{n,s}$ ($s = 1, 2, \ldots, S$) from the posterior distribution for each event $n$; that is, each $x_{n,s}$ is drawn from a distribution proportional to $p(d_n|x_n)p_{\rm ref}(x_n)$. This approach captures degeneracies in the posterior not previously elaborated in detail, most notably the well-known strong correlations between the inferred binary's component masses and spins (e.g., between $\chi_{\rm eff}$ and $m_2/m_1$).[2] Equivalently, this approach gives us direct access to properties of the posterior distribution that were not reported in published tables [9], most notably for the relative posterior probabilities of different choices for binary BH spins (e.g., the data underlying Figure 2.3).

### 2.3.2 Comparing models to observations

The overall likelihood of GW data $\{d\}$ using a model parameterized by $\Lambda$ is [87]

$$p(\{d\}|\Lambda) \propto e^{-\mu} \prod_n \int \mathrm{d}x_n \, p(d_n|x_n) \, \mathcal{R} \, p(x_n|\Lambda) \tag{2.2}$$

where $x_n$ denote candidate intrinsic and extrinsic parameters for the $n$th observation, $\mu$ is the expected number of detections according to the formation scenario $\Lambda$, $p(d_n|x_n)$ is the likelihood for event $n$; $p(\{d\}|\Lambda)$ is the marginalized likelihood; $p(x_n|\Lambda)$ is the prior evaluated at event $n$; and $\mathcal{R}$ (implicitly depending on $\Lambda$ as well) is the average number of merger events per unit time and volume in the Universe. In this expression, we have subdivided the data $\{d\}$ into data with confident detections $d_1, d_2, \ldots, d_N$ and the remaining data; the Poisson prefactor $\exp(-\mu)$ accounts for the absence of detections in the remaining data; and the last product accounts for each independent observation $d_n$. Combined, the factors $e^{-\mu} \prod_n \mathcal{R} p(x_n)$ are the distribution function for an inhomogeneous Poisson process used to characterize the formation and detection of coalescing BH binaries [88, 89]. As described in Appendix B.2, the probability density functions $p(x|\Lambda)$ are estimated from the weighted samples that define each synthetic universe $\Lambda$, and the integrals $\int p(d|x)p(x|\Lambda)$ are performed efficiently via Monte Carlo integration. Similarly, the expected number of detections $\mu$ at O1 sensitivity – a known constant for each model $\Lambda$ – is already provided by the detailed cosmological integration performed in prior work; see Sec. 2.2 and Figure 2.1. Since the marginal likelihood can always be evaluated, the model inference on our discrete set of models becomes an application of Bayesian statistics. In this work, we report the Bayes factor or likelihood ratio $K_{ij} = p(\{d\}|\Lambda_i)/p(\{d\}|\Lambda_j)$ between two different sets of assumptions. To fix the zero point for the log Bayes factor, we adopt the M16 model with $\chi_1 = \chi_2 = 0.5$, henceforth denoted collectively as $J$, and henceforth use $\ln K$ as shorthand for $\ln K_{iJ}$.

[2] Different properties of the binary, like the masses and spins, influence the inspiral, and thus the radiation $h(t)$, in generally different ways; however, sometimes, several parameters can influence the radiation in a similar or degenerate way. For example, both the binary mass ratio and (aligned) binary spin can extend the duration of the inspiral. Similarly, both the binary masses and spins – 8 parameters – determine the final complex frequency of the BH – at leading order, only set by two parameters. Due in part to degeneracies like these, LIGO's inferences about the parameters $x$ for each merging BH lead to a highly correlated likelihood $p(d|x)$ and hence posterior distribution; see, e.g. [12, 55, 56] and references therein.

In what follows, we will mainly discuss comparisons of our models to all of LIGO's reported detection candidates in O1: GW150914, GW151226, and LVT151012 [9]. We do this because LIGO's O1 observational time and survey results are well-defined and comprehensively reported [9]; because we can employ detailed inference results for all O1 events; and because, as we show below, adding GW170104 to our analysis produces little change to our results. Using the approximate posterior described in Appendix B.3 for GW170104, we will also compare all reported LIGO observations (O1 and GW170104) to our models.

Critically, for clarity and to emphasize the information content of the data, in several of our figures we will illustrate the marginal likelihood of the data $p(\{d\}|\Lambda)$ evaluated assuming all binaries are formed with *identical* natal spins. These strong assumptions in our illustrations show just how much the data informs our understanding of BH natal spins. With only four observations, assumptions about the spin distribution are critical to make progress. As described in Appendix B.2, we can alternatively evaluate the marginal likelihood accounting for any concrete spin distribution, or even all possible spin distributions – in our context, all possible mixture combinations of the 100 different choices for $\chi_1$ and $\chi_2$ that we explored. In the latter case, as we show below, just as one expects a priori, observations cannot significantly inform this 100-dimensional posterior spin distribution. As suggested in previous studies [e.g. 14, 56, 60, 66], LIGO's observations in O1 and O2 can be fit by models that includes a wide range of progenitor spins, so long as sufficient probability exists for small natal spin and/or significant misalignment. As a balance between complete generality on the one hand (a 100-dimensional distribution of natal spin distributions) and implausibly rigid assumptions on the other (fixed natal spins), we emphasize a simple one-parameter model, where BH natal spins

$\chi$ are drawn from the piecewise constant distribution

$$p(\chi) = \begin{cases} \lambda_A/0.6 & \chi \leq 0.6 \\ (1-\lambda_A)/0.4 & 0.6 < \chi < 1 \end{cases} \tag{2.3}$$

where $\lambda_A$ is the probability of a natal spin $\leq 0.6$ and the choice of cutoff 0.6 is motivated by our results below.

## 2.4 Results

In this section we calculate the Bayes factor $\ln K$ for each of the binary evolution models described above. Unless otherwise noted, we compare our models to LIGO's O1 observations (i.e., the observation of GW150914, GW151226, and LVT151012), using each model's correlated predictions for the event rate, joint mass distribution $(m_1, m_2)$, $\chi_{\rm eff}$ distribution, and the distribution of $\theta_1, \theta_2$. For numerical context, a Bayes factor of $\ln 10 \simeq 2.3$ is by definition equivalent to 10:1 odds in favor of some model over our reference model. Bayes factors that are more than 5 below the largest Bayes factor observed are in effect implausible (e.g., more than 148:1 odds against), whereas anything within 2 of the peak are reasonably likely.

### 2.4.1 Standard scenario and limits on BH natal spins (O1)

The M10 model allows us to examine the implications of binary evolution with effectively zero natal kicks. The M10 model adopts fiducial assumptions about binary evolution and BH natal kicks, as described in prior work [34, 62]. In this model, BH kicks are suppressed by fallback; as a result, the heaviest BHs receive nearly or exactly zero natal kicks and hence have nearly or exactly zero spin-orbit misalignment.

If heavy BH binaries have negligible spin-orbit misalignment, then natal BH spins are directly constrained from LIGO's measurements (e.g., of $\chi_{\rm eff}$). For example, LIGO's observations of GW150914 severely constrain its component spins to be small, if the spins must be strictly and positively aligned [55, 56]. Conversely, however, LIGO's observations for GW151226 re-quire some nonzero spin. Combined, if we assume all BHs have spins drawn from the same, mass-independent distribution and have negligible spin-orbit misalignments, then we conclude BH natal spins should be preferentially small. [We will return to this statement in Section 2.4.4.]

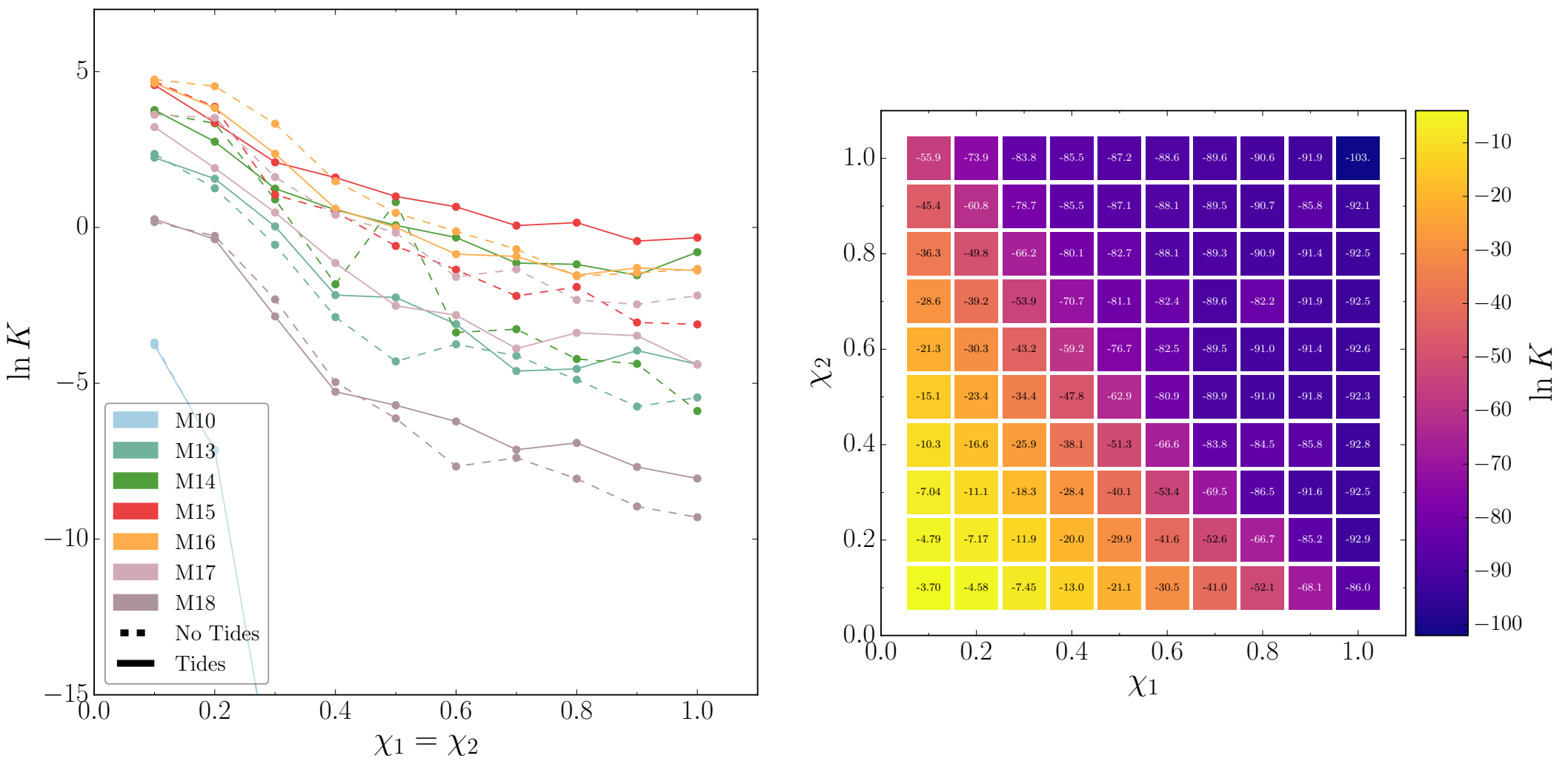

Figure 2.4: **Standard small-kick scenario (M10) requires small natal BH spin**: *Left panel*: A plot of the Bayes factor $K$ derived by comparing GW151226, GW150914, and LVT151012 to the M10 (blue) formation scenario, versus the magnitude of assumed BH natal spin $\chi_1 = \chi_2$. All other models are shown for comparison. Colors and numbers indicate the Bayes factor; dark colors denote particularly unlikely configurations. *Right panel*: As before (i.e., for M10), but in two dimensions, allowing the BH natal spins for the primary and secondary BH to be independently selected (but fixed); color indicates the Bayes factor. As this scenario predicts strictly aligned spins for the heaviest BH-BH binaries, only small BH natal spins are consistent with LIGO's constraints on the (aligned) BH spin parameter $\chi_{\rm eff}$ in O1 (and GW170104); see **(author?)** [90, 3] and [60].

Figure 2.4 shows one way to quantify this effect within the context of our calculations. The left panel shows the Bayes factor for all of our formation models (including M10) as a function of BH natal spin, assuming all BHs have the same (fixed) natal spin $\chi = \chi_1 = \chi_2$. As expected from LIGO's data, large natal BH spins cannot be adopted with M10 and remain consistent with LIGO's observations. The right panel shows the Bayes factor for M10 as function of both BH natal spins, allowing the more massive and less massive BHs to receive different (fixed) natal spins. [The blue line on the left panel uses precisely the same data as the diagonal $\chi_1 = \chi_2$ on the right.] The colorscale graphically illustrates the same conclusion:

though marginally greater freedom exists for natal BH spin on the smaller of the two BHs, we can rule out that all BHs, independent of their mass, have significant natal spin if M10 is true. Conversely, if M10 is true and all BHs have the same natal spins, then this natal spin is likely small.

### 2.4.2 BH natal kicks and misalignment (O1)

In the absence of BH natal kicks, the preponderance of observed BH-BH binaries consistent with $\chi_{\rm eff} \simeq 0$ (e.g., GW150914 and GW170104) provided *conditional* evidence in favor of small BH natal spins. But even small BH natal kicks can frequently produce significant spin-orbit misalignment. Once one incorporates models that permit nonzero BH natal kicks, then even binary BHs with large BH natal spins could be easily reconciled with every one of LIGO's observations. Figures 2.4 and 2.5 provide a quantitative illustration of just how much more easily models with even modest BH natal kicks can explain the data, for a wide range of BH spins. When natal kicks greater than 25km/s are included, the BH natal spin is nearly unconstrained. As is particularly apparent in Figure 2.5, some natal BH spin is required to reproduce the nonzero spin seen in GW151226.

Larger kicks produce frequent, large spin-orbit misalignments and therefore greater consistency with the properties of all of LIGO's observed binary BHs. Spin-orbit misalignment is consistent with the spin distribution of GW151226, and helpful to explain the distribution of $\chi_{\rm eff}$ for LIGO's other observations. However, larger kicks also disrupt more binaries, substantially decreasing the overall event rate (see Figure 2.1). Figure 2.5 illustrates the tradeoff between spin-orbit misalignment and event rate.

### 2.4.3 Tides and realignment (O1)

All other things being equal, our "no tides" scenarios most frequently produce significant spin-orbit misalignment. As a result, even for large BH natal spins, these models have a greater ability to explain LIGO's observations, which are largely consistent with $\chi_{\rm eff} \simeq 0$. The "tides" scenario produces smaller misalignments for the second-born BH. Figure 2.5 quantitatively illustrates how the "no tides" scenario marginally fits the data better. In order to reproduce the inferred distribution of spin-orbit misalignments (in GW151226) and low $\chi_{\rm eff}$ (for all events so far), the "tides" models likely have (a) larger BH natal kicks $\simeq$ 200kms/s and (b) low BH natal spins $\chi_{1,2} \lesssim 0.2$. Conversely, when "no tides" act to realign the second BH spin, *small* natal kicks $\simeq$ 50km/s are favored. Figure 2.5 illustrates the two distinct conclusions about BH natal kick strength drawn, depending on whether stellar tidal realignment is efficient or inefficient. Based on this figure (and hence on the assumption of fixed natal spins), we estimate that massive BHs should receive a natal kick of $\sim$ 50 km/s if no processes act to realign stellar spins. Significantly larger natal kicks, with one-dimensional velocity dispersion $\simeq$ 200km/s, will be required if stellar spins efficiently realign prior to the second BH's birth.

Tides also introduce an asymmetry between the spin-orbit misalignment of the first-born (generally more massive) and second-born (generally less massive) BH [38]. As a result, when we consider general prescriptions for BH natal spins $\chi_1 \neq \chi_2$, we find that scenarios without tides produce largely symmetric constraints on $\chi_{1,2}$. When we assume tidal alignment, we can draw stronger constraints about the second-born spin rather than the first. Paradoxically, large natal spin on the *first* born BH is consistent with observations. The second born BH cannot significantly misalign its spin through a natal kick; therefore, for comparable mass binaries like GW150914, we know that the second-born BH spin must be small, if it is strongly aligned. More broadly, since observations rule out large $\chi_{\rm eff}$, binary formation scenarios with tides and with $\chi_1 > \chi_2$ fit the data substantially better than scenarios with tides and $\chi_2 > \chi_1$. Because tides act to realign the second spin, only when $\chi_2 \leq \chi_1$ will we have a chance at producing small $|\chi_{\rm eff}|$, as LIGO's O1 observations suggest. Figure 2.6 illustrates this asymmetry.

The illustrative results described in this section follow from our strong prior assumptions: fixed BH natal spins. As described below, if we instead adopt some broad distribution of BH natal spins, the substantially greater freedom to reproduce LIGO's observations reduces our ability to draw other distinctions, in direct proportion to the complexity of the prior hypotheses explored. We describe results with more generic spin distributions below.

### 2.4.4 BH natal spins, given misalignment (O1)

So far, to emphasize the information content in the data, we have adopted the simplifying assumption that each pair of BHs has the same natal spins $\chi_1, \chi_2$. This extremely strong family of assumptions allows us to leverage all four observations, producing large changes in Bayes factor as we change our assumptions about (all) BH natal spins. Conversely, if the BH natal spins are nondeterministic, drawn from a distribution with support for any spin between 0 and 1, then manifestly only four observations cannot hope to constrain the BH natal spin distribution, even were LIGO's measurements to be perfectly informative about each BH's properties. Astrophysically-motivated or data-driven prior assumptions must be adopted in order to draw stronger conclusions about BH spins (cf. [91]).

As a concrete example, we consider the simple two-bin BH natal spin model described in Eq. (2.3), with probability $\lambda_A$ that any BH has natal spin $\chi_i \leq 0.6$ and probability $1-\lambda_A$ that any BH natal spin is larger than 0.6. The choice of 0.6 is motivated by our previous results in Figure 2.5, as well as by the empirical $\chi_{\rm eff}$ distribution shown in Figure 2.3. Using the techniques described in Appendix B.2, we can evaluate the posterior probability for $\lambda_A$ given LIGO's O1 observations, within the context of each of our binary evolution models. Figure 2.7 shows the result: LIGO's observations weakly favor low BH natal spins. For models like M10 and M13, with minimal BH natal kicks and hence spin-orbit misalignment, low BH natal spin is necessary to reconcile models with the fact that LIGO hasn't seen BH-BH binaries with large, aligned spins and thus large $\chi_{\rm eff}$. Conversely, LIGO's observations will modestly less strongly disfavor models that frequently predict large BH natal spins (e.g., $\lambda \lesssim 0.6$).

As we increase the complexity of our prior assumptions, our ability to draw conclusions from only four observations rapidly decreases. For example, we can construct the posterior distribution for a generic BH natal spin distribution (i.e., our mixture coefficients $\lambda_\alpha$ for each spin combination can take on any value whatsoever). The mean spin distribution can be evaluated using closed-form expressions provided in Appendix B.2. In this extreme case, the posterior distribution closely resembles the prior for almost all models, except M10.

To facilitate exploration of alternative assumptions about natal spins and kicks, we have made publicly available all of the marginalized likelihoods evaluated in this work, as supplementary material.

### 2.4.5 Information provided by GW170104

The observation of GW170104 enables us to modestly sharpen all of the conclusions drawn above, due to the reported limits on $\chi_{\mathrm{eff}}$: between $-0.42$ and $0.09$ [3]. Of course, the reported limits for all events must always be taken in context, as they are inferred using very specific assumptions – a priori uniform spin magnitudes, isotropically oriented. Necessarily, inference performed in the context of any astrophysical model for natal BH spins and kicks will draw different conclusions about the allowed range, since the choice of prior influences the posterior spin distribution (see, e.g., [92, 91]). Even taking these limits at face value, however, this one observation can easily be explained using some combination of two effects: a significant probability for small natal BH spins, or some BH natal kicks. First and most self evidently, if all BHs have similar natal spins, then binary evolution models that assume alignment at birth; do not include processes that can misalign heavy BH spins, like M10; and which adopt a common natal BH spin for all BHs are difficult to reconcile with LIGO's observations. On the one hand, GW170104 would require extremely small natal spins in this scenario; on the other, GW151226 requires nonzero spin. Of course, a probabilistic (mixture) model allowing for a wide range of mass-independent BH natal spins can easily reproduce LIGO's observations, even without permitting any alignment; see also [50], which adopts a deterministic model that also matches these two events. Second, binary evolution models with significant BH natal kicks can also explain LIGO's observations. As seen in the bottom left panel of Figure 2.5, large BH natal spins are harder to reconcile with LIGO's observations, if we assume BH spin alignments are only influenced by isotropic BH natal kicks. This conclusion follows from the modest $\chi_{\mathrm{eff}}$ seen so far for all events. Conversely, if we assume efficient alignment of the second-born BH, then the observed distribution of $\chi_{\mathrm{eff}}$ (and $\theta_1$, mostly for GW151226) suggest large BH natal kicks, as illustrated by the bottom right panel of Figure 2.5.

### 2.4.6 Information provided by the mass distribution

The underlying mass distributions predicted by our formation models do depend on our assumptions about BH natal kicks, as shown concretely in Figure 2.8. These modest differences accumulate as BH natal kicks increasingly disrupt and deplete all BH-BH binaries. To quantify the similarity between our distributions, Table 2.1 reports an information-theory-based metric (the KL divergence) that attempts to quantify the information rate or "channel capacity" by which the universe communicates information about the mass distribution to us. If $p(x), q(x)$ are two probability distributions over a parameter $x$, then in general the KL divergence has the form

$$D_{KL}(p|q) = \int dx p(x) \ln[p(x)/q(x)] \tag{2.4}$$

Except for the strongest BH natal kicks, we find our mass distributions are nearly identical. Even with perfect mass measurement accuracy, we would need $\mathcal{O}(1/D_{KL})$ fair draws from our distribution to confidently distinguish between them. As demonstrated by previous studies [88, 93], LIGO will be relatively inefficient at discriminating between the different detected mass distributions. LIGO is most sensitive to the heaviest BHs, which dominate the astrophysically observed population, but has extremely large measurement uncertainty in this regime. Thus, accounting for selection bias and smoothing using estimated measurement error, the mass distributions considered here look fairly similar [88]. For constraints on BH natal kicks, the information provided by the mass distribution is far less informative than the insights implied by constraints on $\chi_{\rm eff}$ and $\theta_{1,2}$.

As a measure of the information LIGO can extract per event about the mass distribution from each detection, we enumerate how many different BH-BH binaries LIGO can distinguish, which are consistent with the expected stellar-mass BH-BH population (i.e., motivated by LIGO's reported observations to date, limiting to $m_2/m_1 > 0.5$, $m_1+m_2 < 75M_\odot$, $m_2 > 3M_\odot$, and $m_1 < 40M_\odot$). Counting up the distinct waveforms used by gravitational wave searches in O2 [94], including spin, there are only **236** templates with chirp masses above LVT151012 (i.e., $\mathcal{M}_c > 15M_\odot$), and only $\simeq$ **1,200** with chirp masses above GW151226 (i.e., $\mathcal{M}_c > 8.88M_\odot$). This estimate is highly optimistic, because it neglects distance and hence redshift uncertainty, which decreases our ability to resolve the smallest masses (i.e., the uncertainty in chirp mass for GW151226), and it also uses both mass and spin information. Judging from the reported mass distributions alone (e.g., the top left panel of Figure 4 in [9]), LIGO may efficiently isolate BHs to only a few tens of distinct mass bins, de facto limiting the resolution of any mass distribution which can be nonparametrically resolved with small-number statistics; see, e.g., the discussion in [93].

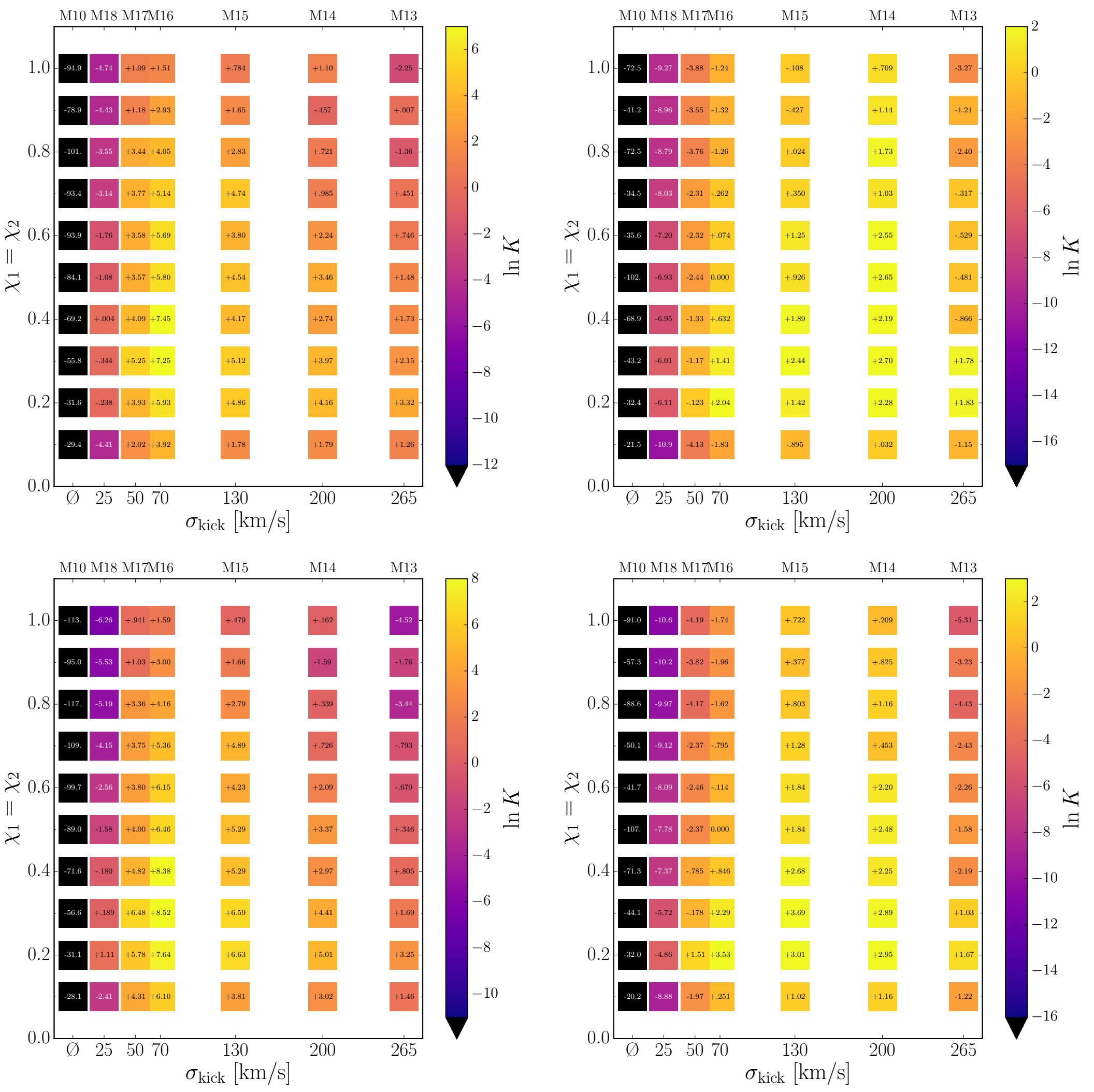

Figure 2.5: **Bayes factor versus spin and kicks, with and without tides:** A plot of the Bayes factor versus BH natal spin ($\chi = \chi_1 = \chi_2$) and natal kick ($\sigma_{\mathrm{kick}}$). The left and right panels correspond to "no tides" and "tides", respectively. The top two panels use only the O1 events; the bottom two panels account for the events and network sensitivity updates reported in the GW170104 discovery paper. In each panel, the zero point of the Bayes factor is normalized to the BH-BH formation scenario with $\chi_1 = \chi_2 = 0.5$ and $\sigma = 70$ km/s and "tides".

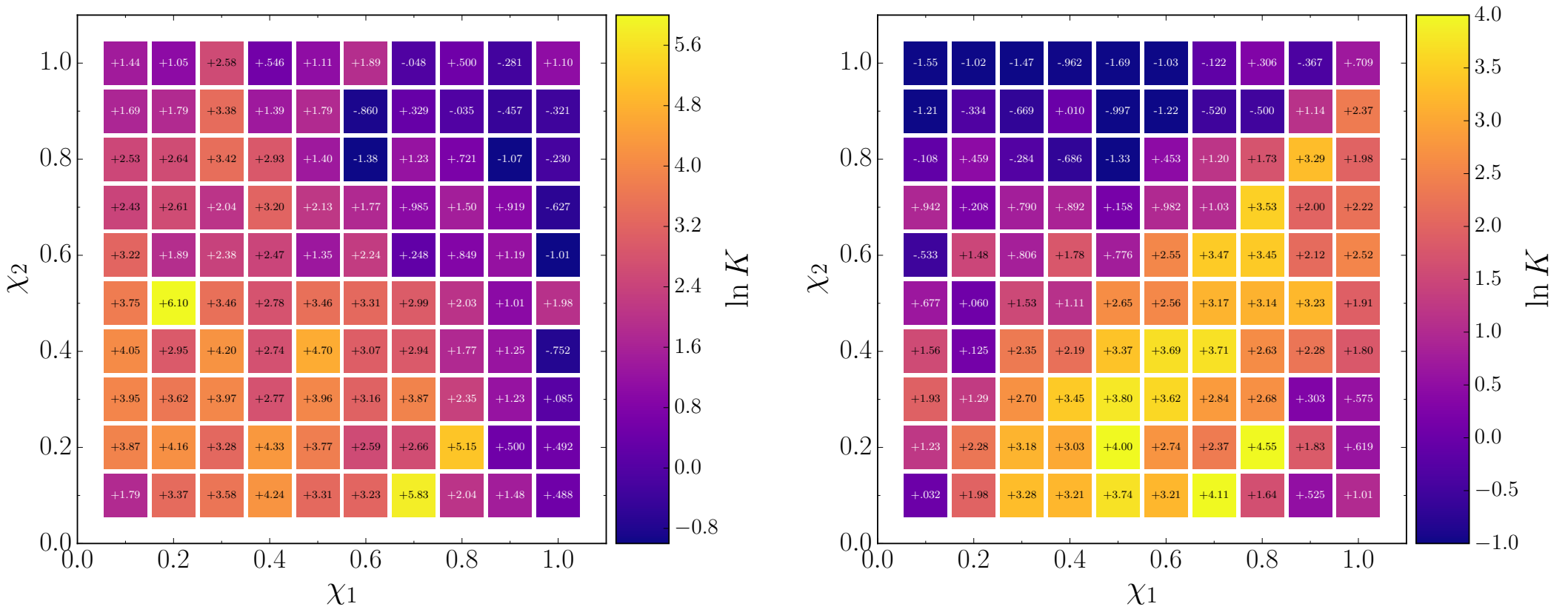

Figure 2.6: **Bayes factor versus spin, with and without tides (O1)**: For the M14 model ($\sigma = 200$km/s), a plot of the Bayes factor versus $\chi_{1,2}$. Colors and numbers indicate the Bayes factor; dark colors denote particularly unlikely configurations. The left panel assumes no spin realignment ("no tides"); the right panel assumes the second-born BH's progenitor had its spin aligned with its orbit just prior to birth ("tides"). Spin-orbit realignment and the high orbital velocity just prior to the second SN ensures the second spin is at best weakly misaligned; therefore, $\chi_2$ would need to be small for these models to be consistent with LIGO's observations to date.

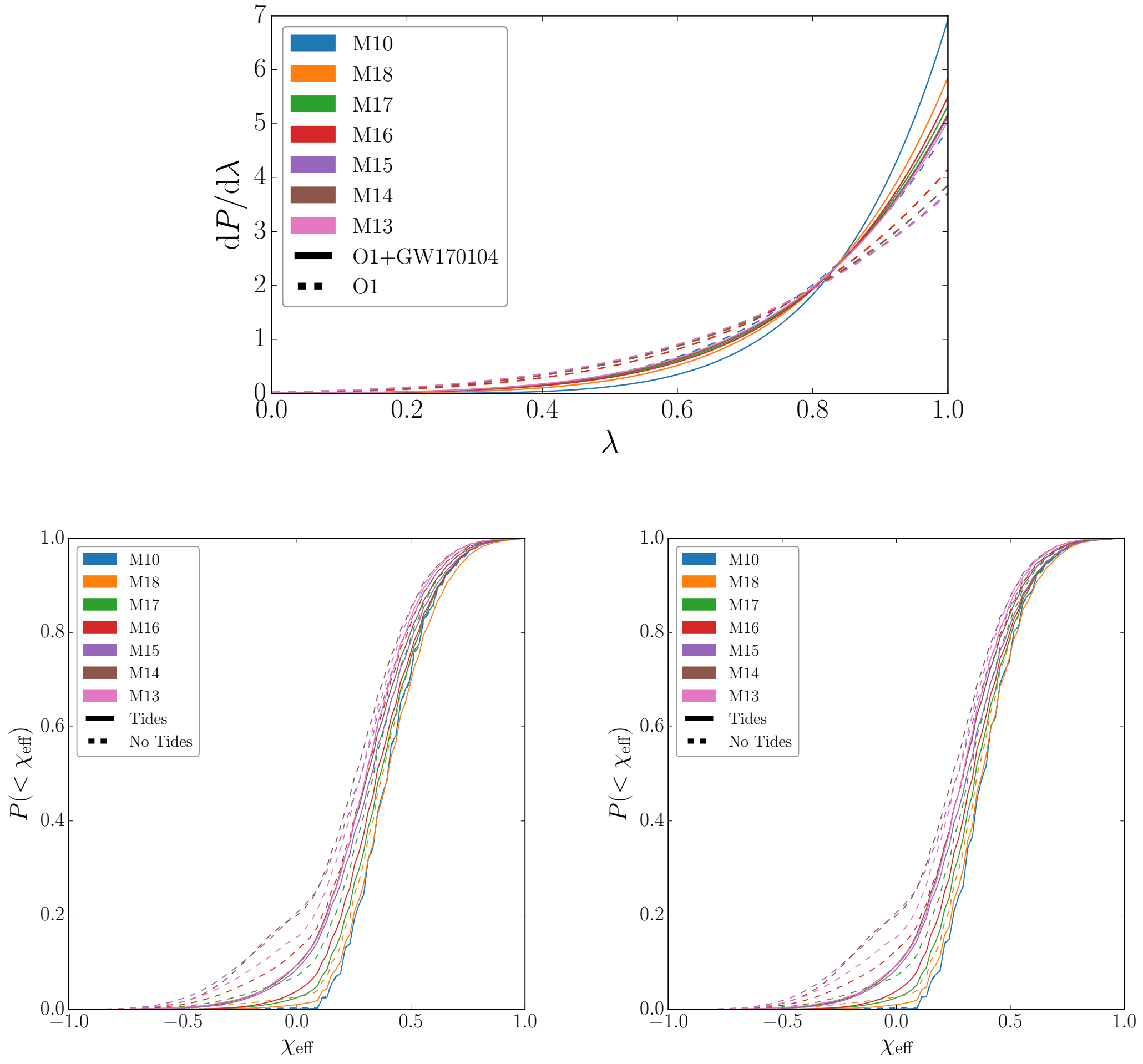

Figure 2.7: **High or low natal spin?** *Top panel*: Posterior distribution on $\lambda_A$, the fraction of BHs with natal spins $\leq 0.6$ [Eq. (2.3)], based on O1 (dotted) or on O1 with GW170104 (solid), compared with our binary evolution models (colors), assuming "no tides". Unlike Figure 2.5, which illustrates Bayes factors calculated assuming fixed BH natal spins, this calculation assumes each BH natal spin is drawn at random from a mass- and formation-scenario-independent distribution that is piecewise constant above and below $\chi = 0.6$. With only four observations, LIGO's observations consistently but weakly favor low BH natal spins. *Left panel*: Posterior distribution for $\chi_{\rm eff}$ implied by the distribution of $\lambda_A$ shown in the top panel (i.e., by comparing our models to LIGO's O1 observations, under the assumptions made in Eq. (2.3)). *Right panel*: As in the left panel, but including GW170104. Adding this event does not appreciably or qualitatively change our conclusions relative to O1.

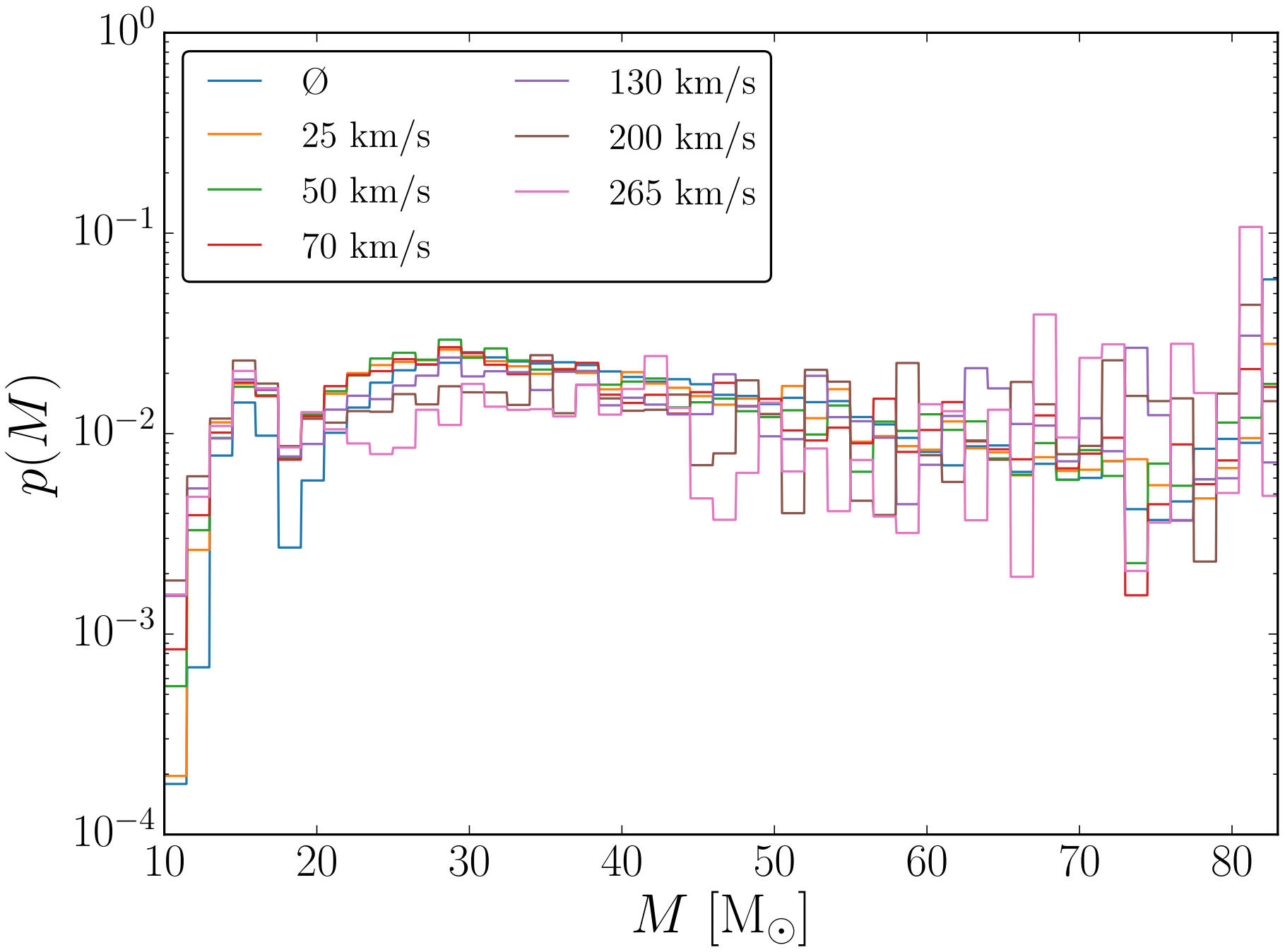

Figure 2.8: Detection-weighted total mass distributions of our models, labeled by their $\sigma_{\mathrm{kick}}$ values, without accounting for LIGO measurement error. The overall mass distributions are very similar, particularly for low kicks.

## 2.5 Predictions and projections

Using the Bayes factors derived above for our binary evolution models and BH natal spin assumptions (collectively indexed by $\Lambda$), we can make predictions about future LIGO observations, characterized by a probability distribution $p_{\rm future}(x) = \sum_\Lambda p(x|\Lambda)p(\Lambda|d)$ for a candidate future binary with parameters $x$. We can then account for LIGO's mass-dependent sensitivity to generate the relative probability of observing binaries with those parameters. In the context of the infrastructure described above, we evaluate this detection-weighted posterior probability using a mixture of synthetic universes, with relative probabilities $p(\Lambda|d)$ and relative weight $r_i$ of detecting an individual binary drawn from it.

Using our fiducial assumptions about BH spin realignment ("no tides"), our posterior probabilities point to nonzero BH natal kicks, with BH natal spins that can neither be too large nor too small (Figures 2.5 and 2.7). In turn, because each of our individual formation scenarios $\Lambda$ preferentially forms binaries with $\chi_{\rm eff} > 0$ [61], with a strong preference for the largest $\chi_{\rm eff}$ allowed, we predict future LIGO observations will frequently include binaries with the largest $\chi_{\rm eff}$ allowed by the BH natal spin distribution. These measurements will self-evidently allow us to constrain the natal spin distribution (e.g., the maximum natal BH spin). For example, if future observations continue to prefer small $\chi_{\rm eff}$, then the data would increasingly require smaller and smaller natal BH spins, within the context of our models. For example, this future scenario would let us rule out models with large kicks and large spins, as then LIGO should nonetheless frequently detect binaries with large $\chi_{\rm eff}$.

As previously noted, with only four GW observations, the data does not strongly favor any spin magnitude distribution. Strongly modeled approaches which assume specific relationships between the relative prior probability of different natal spins can draw sharper constraints, as in [60]. If we allow the spin distribution to take on any form [7, 95], many observations would be required to draw conclusions about the spin distribution. Conversely, as described previously and illustrated by Figure 2.7, if we adopt a weak (piecewise-constant) model, we can draw some weak conclusions about the BH natal spin distributions that are implied by our binary evolution calculations and LIGO's observations.

Neither the expected number of events nor their mass distribution merits extensive discussion. The large Poisson error implied by only four observations leads to a wide range of probable event rates, previously shown to be consistent with all the binary evolution models presented here [34, 62]. Conversely, due to the limited size of our model space – the discrete model set and single model parameter (BH natal kicks) explored – these posterior distributions by no means fully encompass all of our prior uncertainty in binary evolution and all we can learn by comparing GW observations with the data. While our calculations illuminate how GW measurements will inform our understanding of BH formation, our calculations are not comprehensive enough to provide authoritative constraints except for the most robust features.

Finally, all of our calculations and projections have been performed in the context of one formation scenario – isolated binary evolution. Globular clusters could also produce a population of merging compact binaries [17], with random spin-orbit misalignments [96]. Several previous studies have described or demonstrated how to identify whether either model contributes to the detected population, and by how much, using constraints on merging BH-BH spins [14, 23, 97, 81, 13, 95].

## 2.6 Conclusions

By comparing binary evolution models with different assumptions about BH natal kicks to LIGO observations of binary BHs, we estimate that heavy BHs should receive a natal kick of order 50 km/s if no processes act to realign stellar spins. Significantly larger natal kicks, with one-dimensional velocity dispersion $\simeq$ 200km/s, will be required if stellar spins efficiently realign prior to the second BH's birth. These estimates are consistent with observations of galactic X-ray binary misalignment [40, 41, 42, 43] and recoil velocity [44, 45, 46, 47, 43, 48, 49]. Our estimate is driven by two simple factors. The natal kick dispersion $\sigma$ is bounded from above because large kicks disrupt too many binaries (reducing the merger rate below the observed value). Conversely, the natal kick distribution is bounded from below because modest kicks are needed to produce a range of spin-orbit misalignments. A distribution of misalignments increases our models' compatibility with LIGO's observations, if all BHs are likely to have natal spins.

Given limited statistics, we have for simplicity (and modulo M10) assumed all binary BHs receive natal kicks and spins drawn from the same formation-channel-independent distributions. This strong assumption about BH natal spins allows us to draw sharp inferences about BH natal spins and kicks by combining complementary information provided by GW151226 (i.e., nonzero spins required, with a suggestion of misalignment) and the remaining LIGO observations (i.e., strong limits on $\chi_{\rm eff}$). Future observations will allow us to directly test more complicated models not explored here, where the natal spin and kick distribution depends on the binary BH mass as in **(author?)** [66] Necessarily, if BH natal spins are small for massive BHs and large for small BHs, as proposed in **(author?)** [66], then measurements of low-mass BH binaries like GW151226 will provide our primary channel into constraining BH natal spins and kicks. At present, however, inferences about BH natal spins and spin-orbit misalignment are strongly model or equivalently prior driven, with sharp conclusions only possible with strong assumptions. We strongly recommend results about future BH-BH observations be reported or interpreted using multiple and astrophysically motivated priors, to minimize confusion about their astrophysical implications (e.g., drawn from the distribution of $\chi_{\rm eff}$).

For simplicity, we have also only adjusted one assumption (BH natal kicks) in our fiducial model for how compact binaries form. A few other pieces of unknown and currently-parameterized physics, notably the physics of common envelope evolution, should play a substantial role in how compact binaries form and, potentially, on BH spin misalignment. Other assumptions have much smaller impact on the event rate and particularly on BH spin misalignment. Adding additional sources of uncertainty will generally diminish the sharpness of our conclusions. For example, the net event rate depends on the assumed initial mass function as well as the star formation history and metallicity distribution throughout the universe; once all systematic uncertainties in these inputs are inclusded, the relationship between our models and the expected number of events is likely to include significant systematic as well as statistical uncertainty. Thus, after marginalizing over all sources of uncertainty, the event rate may not be as strongly discriminating between formation scenarios. By employing sev-eral independent observables (rate, masses, spins and misalignments), each providing weak constraints about BH natal kicks, we protect our conclusions against systematic errors in the event rate. Further investigations are needed to more fully assess sources of systematic error and enable more precise constraints.

Due to the limited size of our model space – the discrete model set and single model parameter (BH natal kicks) explored – these posterior distributions by no means fully encompass all of our prior uncertainty in binary evolution and all we can learn by comparing GW observations with the data. As in previous early work [98, 99, 100, 101], a fair comparison must broadly explore many more elements of uncertain physics in binary evolution, like mass transfer and stellar winds. Nonetheless, this nontrivial example of astrophysical inference shows how we can learn about astrophysical models via simultaneously comparing GW measurements of several parameters of several detected binary BHs to predictions of any model(s). While we have applied our statistical techniques to isolated binary evolution, these tools can be applied to generic formation scenarios, including homogeneous chemical evolution; dynamical formation in globular clusters or AGN disks; or even primordial binary BHs.

Forthcoming high-precision astrometry and radial velocity from GAIA will enable higher-precision constraints on existing X-ray binary proper motions and distances [102, 103], as well as increasing the sample size of available BH binaries. These forthcoming improved constraints on BH binary velocities will provide a complementary avenue to constrain BH natal kicks using binaries in our own galaxy.

# Chapter 3

# Method II: Bayesian parametric population models

While comparing observations directly with physical population models – as we did in Chapter 2 – is a good way to test existing models, it does not allow for too many surprises. An alternative approach, which serves to complement the other, is to use a purely statistical model for the population's distribution. As an example, if one were given a number of samples from some population, e.g., the lifespans of 10 randomly chosen fish in the Genessee river, one might pick some family of distributions (e.g., a Gaussian) and construct a posterior distribution on the parameters of that distribution (in the case of a Gaussian: $\mu$ and $\sigma$). Of course, the choice of distribution family was somewhat arbitrary, and upon comparison to other families of distributions, it may turn out to be a bad choice. In either case, this approach allows one to answer the question "Given that the population is a member of this family of distributions, which members of that family are most consistent with the data?" When the number of samples grows very large (10 fish are a very small sample to describe the population of a river, but maybe 1000 would do a good job) restricting ourselves like this ceases to be necessary, and more flexible methods (as we describe in Chapter 4) become preferable.

With 3.9 binary black hole detections from LIGO [2, 3], we are in a similar situation to using 10 fish to describe the Genessee river. So at this point, it is useful to consider simple families of distributions which can be parameterized by a few variables. We will develop the general formalism in Section 3.1, then apply it to a power law mass distribution model in Section 3.2, as was done by LIGO [9, 3]. Finally, in Section 3.3, we will discuss some simple extensions that can be made to this model.

## 3.1 General formalism

When estimating the parameters of a population's distribution, there are three kinds of spaces we must keep in mind. There is the physical space-time in which the events occur, the sample space from which each event's parameters are drawn (which we'll denote generically as "$\lambda$"), and the parameter space in which a single point represents one possible population (which we'll denote by a capital "$\Lambda$"). To be completely general, the physical point in space-time of each event should be considered part of the sample space, $\lambda$, as the Universe changes with time, and small-scale anisotropies exist. However, for the remainder of this work, we will assume that the distribution of compact binaries is constant in co-moving volume, and save a more general treatment for later work. As a result of this simplifying assumption, any integrals over space and time will turn into multiplication by the total space and time volume.

Ultimately, what we are interested in, is the population rate density function,

$$\rho(\lambda \mid \Lambda) = \frac{\mathrm{d}N}{\mathrm{d}V\,\mathrm{d}t\,\mathrm{d}\lambda}(\lambda, \Lambda). \tag{3.1}$$

This describes the distribution of event parameters $\lambda$, scaled by an overall rate, depending on the particular population chosen, which is parameterized by $\Lambda$. The utility of this distribution, is that one can obtain the expected number of observations in a region of sample-space, $\Delta\lambda$, and of space and time, $(\Delta V)(\Delta T)$, by integrating over that region

$$\mu_{\mathrm{intrinsic}}(\Lambda) = (\Delta V)(\Delta T) \int_{\Delta\lambda} \rho(\lambda \mid \Lambda)\mathrm{d}\lambda. \tag{3.2}$$

As one would expect, this number is a function of the population, and therefore a function of $\Lambda$. We labeled this $\mu_{\mathrm{intrinsic}}$ to distinguish it from the observed average, which is not the same.

Gravitational wave detectors like LIGO are sensitive to a certain volume of space, and the length of time is dependent only on the effective observing time. However, this volume varies based on the parameters of the system of interest. For instance, more massive black hole binaries produce stronger gravitational waves, and therefore they can be detected at farther distances, making $\Delta V$ larger than it would be for a less massive binary. The binary masses also affect the frequency of the gravitational wave, and since LIGO has only a limited bandwidth, this means there is an upper and lower limit on the masses for which $\Delta V \neq 0$. In addition, the orientation of each binary relative to the detector network will further impact our sensitivity, and therefore it makes sense to instead speak in terms of orientation-averaged volumes, $\langle V \rangle$. We will refer to the combined average sensitive space and time volume as $\langle VT \rangle(\lambda)$, and throughout this work, we will use the same approximation used in [2], which assumes all binaries are non-spinning, and thus $\langle VT \rangle$ depends only on the masses ($m_1$, $m_2$). For a more complete treatment, see Appendix A.2. With all of this in mind, the expected number of observed events in some region of sample-space $\Delta\lambda$ is given by

$$\mu_{\text{observed}}(\Lambda) = \int_{\Delta\lambda} \langle VT \rangle(\lambda) \rho(\lambda \mid \Lambda) \mathrm{d}\lambda. \tag{3.3}$$

To determine the parameters $\Lambda$ of the population's distribution $\rho(\lambda \mid \Lambda)$, we make use of Bayesian inference, which we give some background on in Appendix A.1. We are dealing with an inhomogeneous Poisson process, and thus the likelihood for the population parameters $\Lambda$ and the parameters of $N$ detections $\lambda_1, \ldots, \lambda_N$ is (based on [87])

$$\mathcal{L}(\lambda_1, \ldots, \lambda_N, \Lambda) \propto e^{-\mu(\Lambda)} \prod_{n=1}^{N} \ell_n(\lambda_n) \rho(\lambda_n \mid \Lambda). \tag{3.4}$$

Here, $\ell_n$ is used to denote the likelihood of an individual event's parameters, as would be used when performing parameter estimation for that one event.

Using Bayes' theorem, Equation A.2, one can compute the probability of each model's parameters through the posterior distribution

$$p(\lambda_1, \ldots, \lambda_N, \Lambda \mid d) \propto \pi_\Lambda(\Lambda) e^{-\mu(\Lambda)} \prod_{n=1}^{N} \ell_n(\lambda_n) \rho(\lambda_n \mid \Lambda). \tag{3.5}$$

As one further note, we will regularly separate the shape of $\rho(\lambda \mid \Lambda)$ and its overall normalization, by writing $\rho(\lambda \mid \Lambda) = \mathcal{R} p(\lambda \mid \Lambda)$, where $p$ here is a proper probability distribution, and $\mathcal{R}$ is obtained by integrating $\rho$ over all $\lambda$'s. $\mathcal{R}$ is the intrinsic merger rate, over the entire sample space. In this sense, $\mathcal{R}$ can be thought of as one of the parameters contained within $\Lambda$, and in fact that is how we perform our inference.

## 3.2 Power law mass distribution

Thus far, we have applied this general approach to a single family of models – a power law mass distribution. This was primarily because it is simple and has been used in previous LIGO publications (e.g., [2]), which allows us to confirm our method is consistent with an established result. However, there is still something new to be learned from our more general method, as we infer the rate density $\rho(\lambda \mid \Lambda) = \mathcal{R} p(\lambda \mid \Lambda)$, whereas previous LIGO results have only inferred the probability distribution $p(\lambda \mid \Lambda)$, and separately estimated $\mathcal{R}$ with a single realization of $p(\lambda \mid \Lambda)$.

In the power law mass distribution, we have $\lambda = (m_1, m_2)$, and $\Lambda = (\mathcal{R}, \alpha)$. We start with an auxiliary model. First we take a power law distribution on $m_1$: $p(m_1 \mid \alpha) \propto m_1^{-\alpha}$. We then assume $p(m_2 \mid m_1) = (m_1 - m_{\text{min}})^{-1}$, i.e., it is uniformly distributed up to $m_1$. The joint distribution for our auxiliary model is then

$$p(m_1, m_2 \mid \alpha) \propto \frac{m_1^{-\alpha}}{m_1 - m_{\text{min}}}. \tag{3.6}$$

Now to obtain the model we're actually interested in, we impose the following conditions on

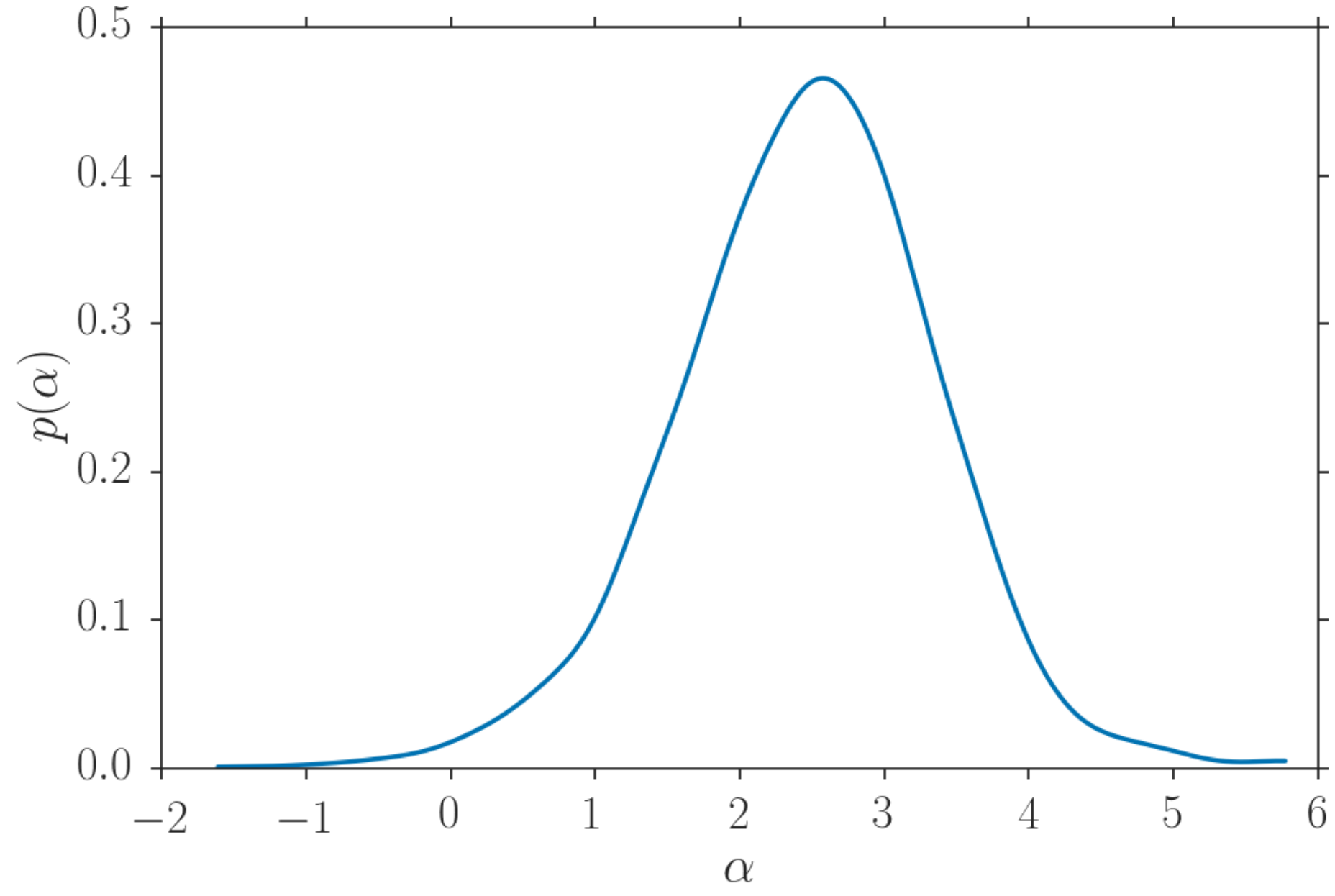

Figure 3.1: Posterior on power law index $\alpha$ for the 2.9 BBH detections in O1. This is nearly identical to the result published in [2], with differences attributable to sampling variance.

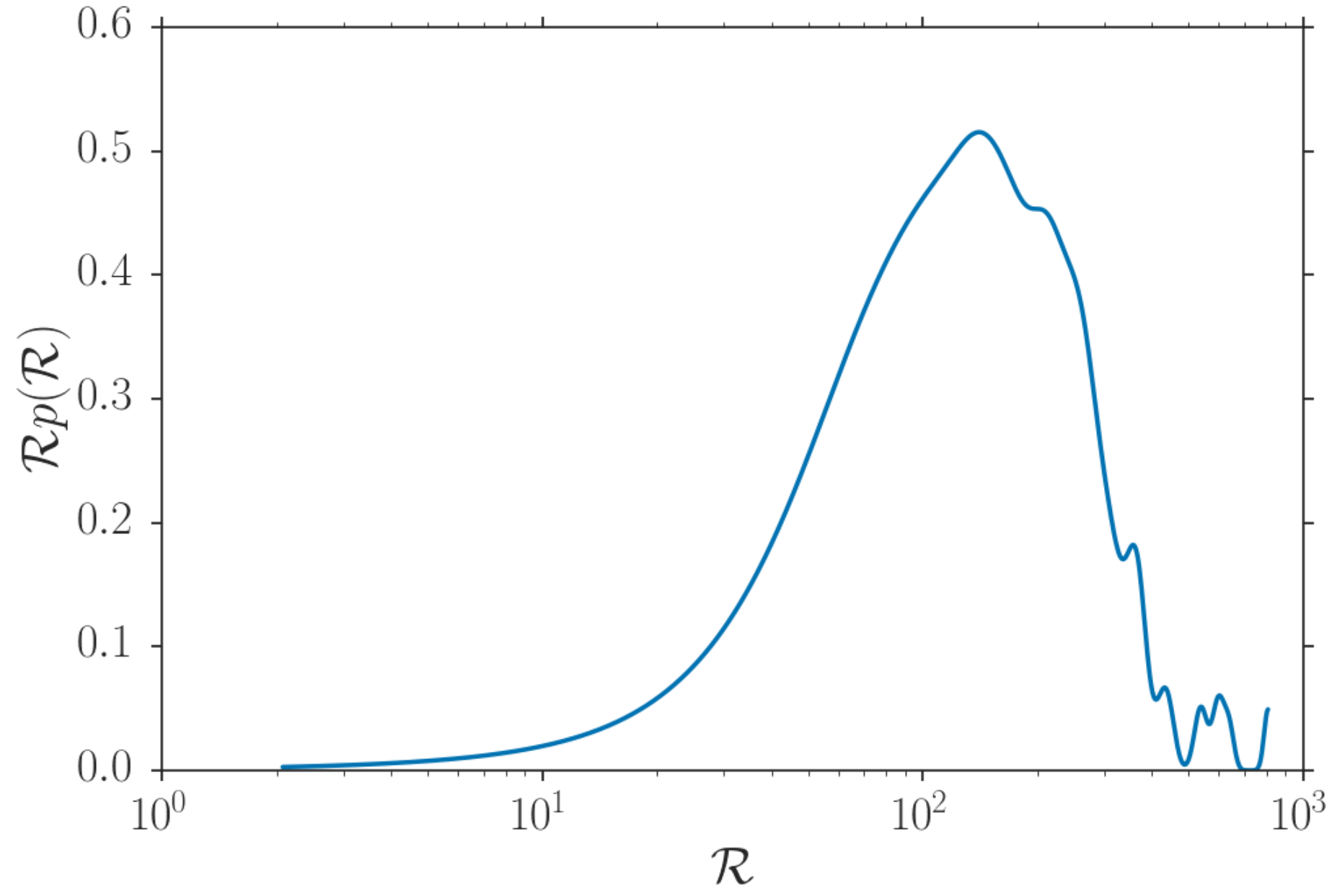

Figure 3.2: Posterior on the rate $\mathcal{R}$ for the 2.9 BBH detections in O1, assuming a power law mass distribution. This differs from the result in [2], which is expected, as the posterior in that paper fixed $\alpha = 2.35$, whereas we allow $\alpha$ to vary. The fluctuations at high $\mathcal{R}$ are attributable to poor sampling at high $\mathcal{R}$, which can be seen in the joint ($\mathcal{R}$, $\alpha$) posterior in Figure 3.3.

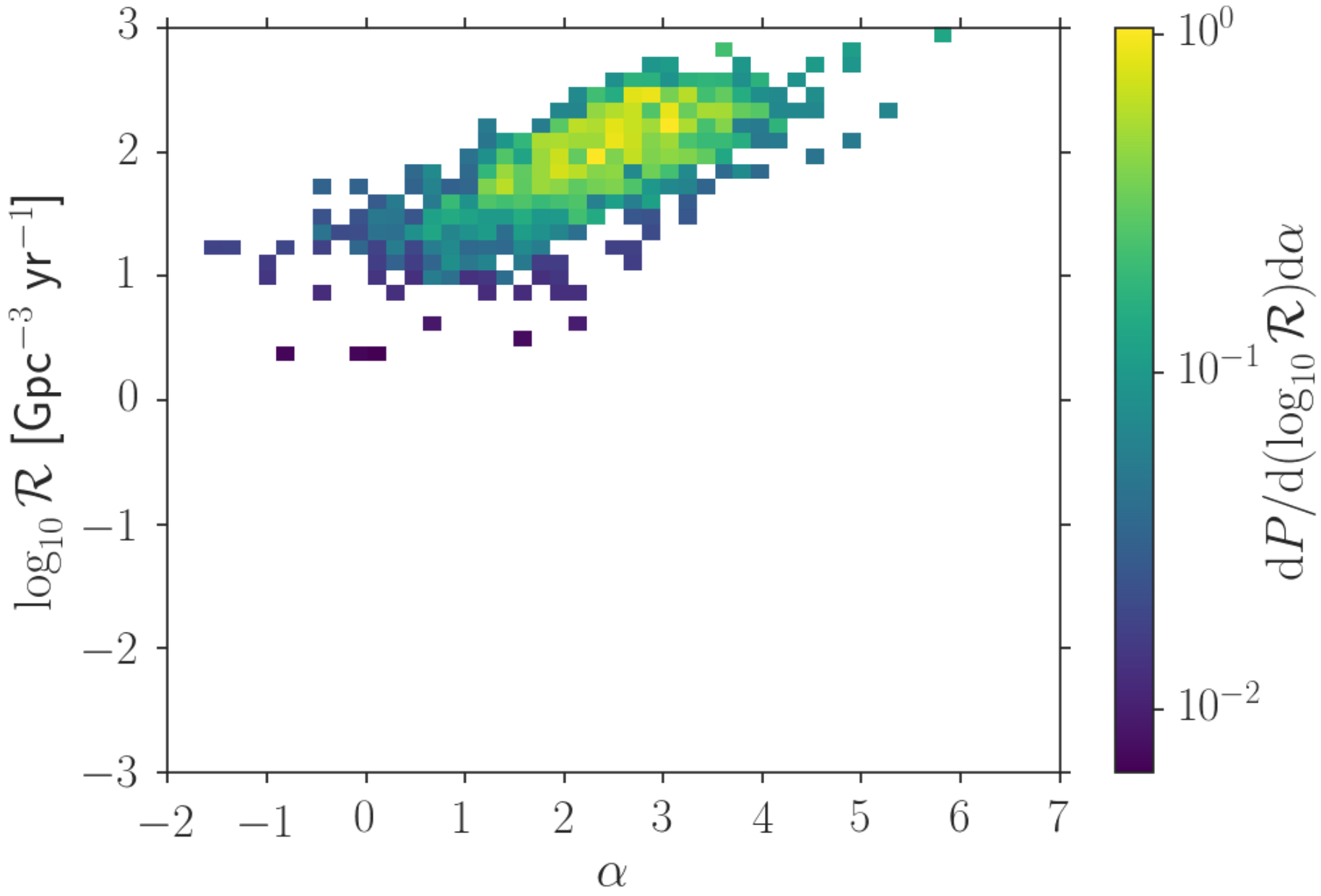

Figure 3.3: Joint posterior on the rate $\mathcal{R}$ (shown in log-scale) and power law index $\alpha$ for the 2.9 BBH detections in O1. This is a new result, as previous power law studies have assumed a fixed $\alpha$ when computing the posterior on $\mathcal{R}$. This is also a taste of what's to come, when we make the other parameters of the model, $m_{\text{min}}$ and $M_{\text{max}}$, free parameters, as that will give us a 4-dimensional joint posterior.

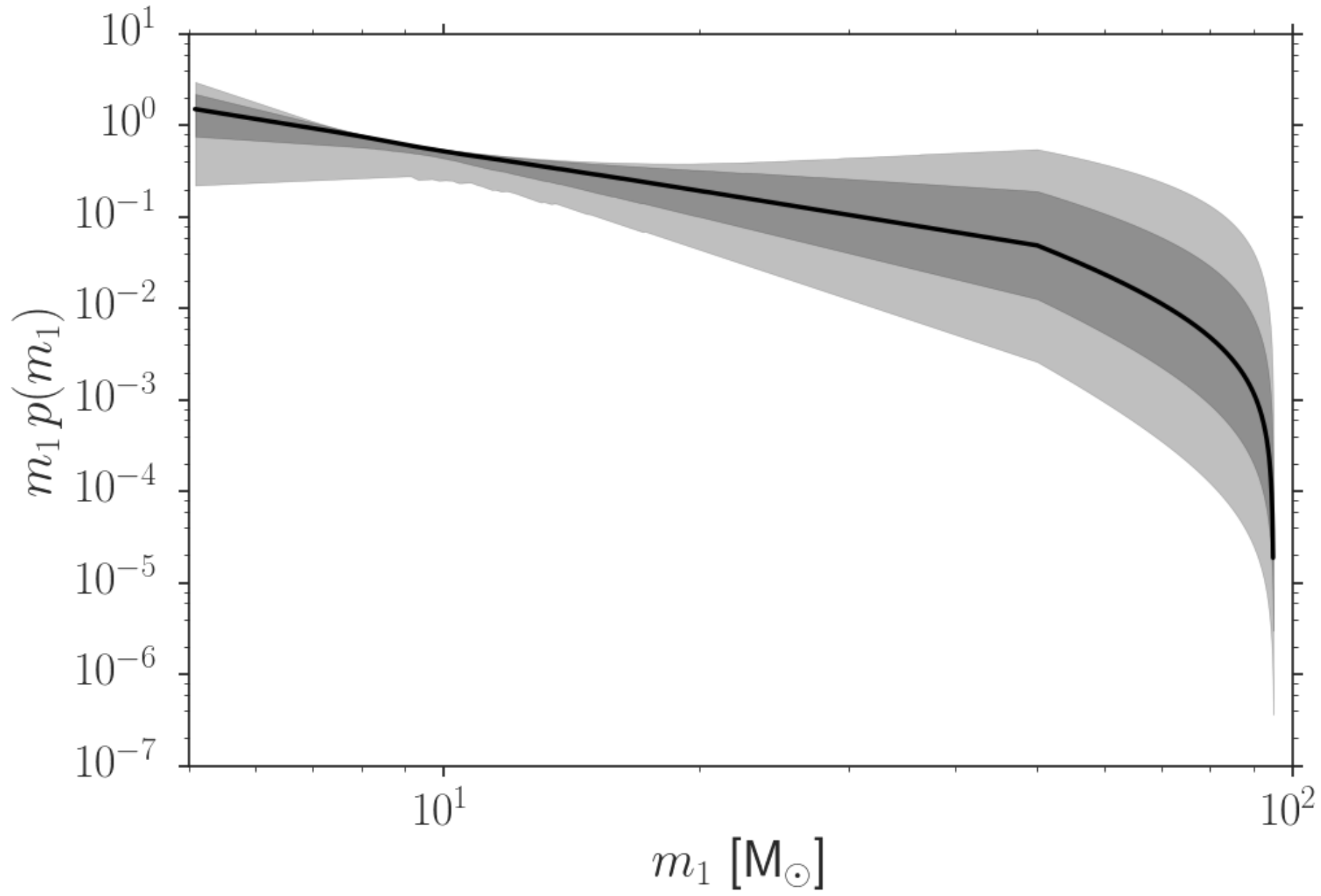

Figure 3.4: The inferred power law mass probability distribution, $p(m_1 \mid \alpha)$, given the 2.9 BBH detections in O1. It has been multiplied by $m_1$ to make it dimensionless. Each value of $\alpha$ results in a different distribution, so what we really have is a distribution of distributions. To visualize this, we show, for each value of $m_1$, the median and $\pm 2\sigma$ credible regions for $p(m_1 \mid \alpha)$.

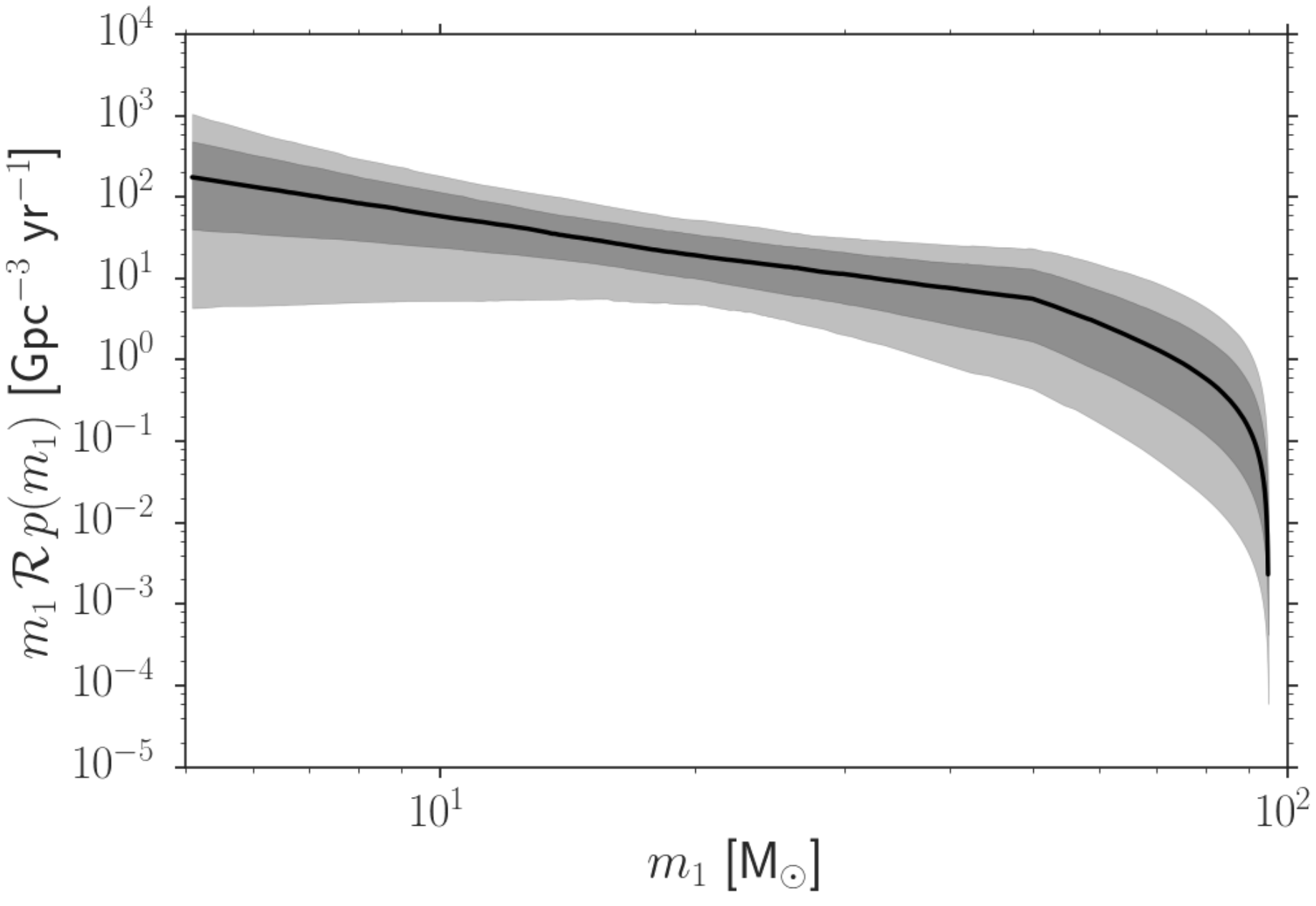

Figure 3.5: The inferred power law mass rate distribution, $\rho(m_1 \mid \alpha)$, given the 2.9 BBH detections in O1. We multiply this by $m_1$ so it only has units of $\mathcal{R}$. Each value of $\mathcal{R}$ and $\alpha$ results in a different distribution, as was the case in Figure 3.4. As before, we show for each value of $m_1$, the median and $\pm 2\sigma$ credible regions for $\rho(m_1 \mid \mathcal{R}, \alpha)$.

$m_1$ and $m_2$:

$$m_1 \geq m_2 \qquad\qquad m_1, m_2 \geq m_{\min} \qquad\qquad m_1 + m_2 \leq M_{\max}. \tag{3.7}$$

Consistent with previous work [2], we choose $m_{\min} = 5\,\mathrm{M}_\odot$ and $M_{\max} = 100\,\mathrm{M}_\odot$.

With these constraints, the marginal distribution on $m_1$ is no longer a power law everywhere, as the joint distribution on $(m_1,\ m_2)$ becomes narrower for certain values of $m_1$ [1]. Integrating out $m_2$, we obtain the marginal distribution

$$p(m_1 \mid \alpha) \propto m_1^{-\alpha} \left( \frac{m_{2,\max} - m_{\min}}{m_1 - m_{\min}} \right), \tag{3.8}$$

where $m_{2,\max} = \min(m_1, M_{\max} - m_1)$, and $m_{\min} < m_1 \leq M_{\max} - m_{\min}$. To get a sense of this distribution and its dependence on $\alpha$, we have plotted it for a few demonstrative values of $\alpha$ in Figure 3.6. For our actual inference, we still make use of Equation 3.6 for $p(\lambda \mid \Lambda)$ – after imposing the constraints on $(m_1, m_2)$.

[1] However, if we take any slice of $p(m_1, m_2 \mid \alpha)$, for a fixed value of $m_2$, that one-dimensional distribution a true power law in $m_1$.

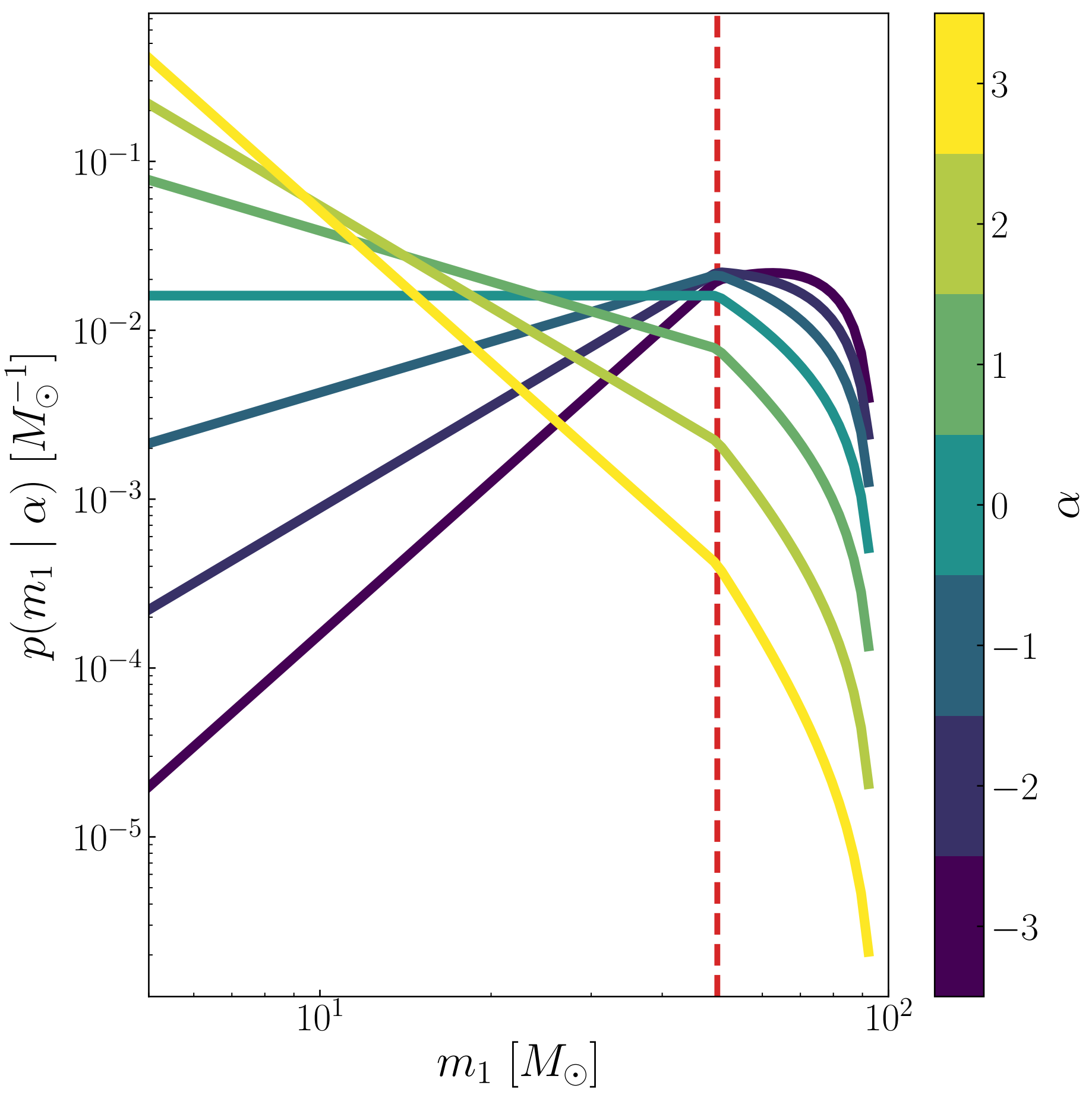

Figure 3.6: Power law mass distribution for several values of $\alpha$. A vertical line has been drawn where $m_1 = M_{\max}/2$ (i.e. $50\,\mathrm{M}_\odot$ in our choice of mass-cuts) as this is the largest mass for which $m_1 = m_2$ is allowed. After this point, $p(m_1 \mid \alpha)$ falls off quickly, as the range of allowed values for $m_2$ begins to shrink.

Now in order to make this a posterior distribution, which we will sample via Markov chain Monte Carlo, we need to introduce a prior on $\Lambda = (\mathcal{R}, \alpha)$. We first assume $\mathcal{R}$ and $\alpha$ are independent, meaning $\pi_\Lambda(\mathcal{R}, \alpha) = \pi_\mathcal{R}(\mathcal{R})\pi_\alpha(\alpha)$. We choose $\pi_\alpha$ to be uniform, conditional on $\alpha_{\min} \leq \alpha \leq \alpha_{\max}$. The precise values of $(\alpha_{\min},\ \alpha_{\max})$ are not important, so long as they contain all of the probable values of $\alpha$, so we chose $(-2,\ 7)$ as the limits, as the posterior on $\alpha$ shown in Figure 12 of [2] – which our results should match exactly – becomes negligible outside that range. Consistent with the rate calculation performed in [2], we used a Jeffrey's prior for $\mathcal{R}$, meaning $\pi_\mathcal{R}(\mathcal{R}) \propto \mathcal{R}^{-1/2}$. We also set bounds $\mathcal{R}_{\min} \leq \mathcal{R} \leq \mathcal{R}_{\max}$, with $(\mathcal{R}_{\min},\ \mathcal{R}_{\max}) = (10^{-3},\ 10^{+3})$, as they include the entire non-negligible region of the posterior. In addition, to make the sampling more efficient, instead of sampling in $\mathcal{R}$, we sampled in $\log \mathcal{R}$, and converted our priors appropriately.

The most important – yet least interesting – result from our method is the $p(\alpha \mid d)$ posterior shown in Figure 3.1. If our method is working properly, then this should be nearly indistinguishable from Figure 12 in [2]. Indeed the differences are barely visible, and may be attributed to MCMC sampling variance, and our event likelihood approximation.

The first original result from our method is the $p(\mathcal{R} \mid d)$ posterior shown in Figure 3.2. Unlike the rate calculation performed in [2], which held $\alpha$ fixed to 2.35, we instead computed the full joint $(\mathcal{R},\ \alpha)$ posterior first, and then marginalized over $\alpha$. This joint posterior is shown in Figure 3.3. It appears that the rate is positively correlated with $\alpha$. This makes sense when we consider what raising $\alpha$ does. As we increase $\alpha$, the relative number of high-mass to low-mass black holes decreases. Since $\langle VT \rangle$ increases with mass, the absence of observations of very low mass black holes is not problematic, as they may be suppressed by selection effects. However, since this drives down the number of high-mass black holes, we must compensate by raising the overall merger rate $\mathcal{R}$. This trade-off balances out in the highest probability region of $p(\mathcal{R}, \alpha \mid d)$, which occurs around $\mathcal{R} \simeq 100\,\mathrm{Gpc}^{-3}\,\mathrm{yr}^{-1}$ and $\alpha \simeq 3$. Going back to the marginal posterior on $\mathcal{R}$, this extra freedom from $\alpha$ has skewed $p(\mathcal{R}|d)$ towards higher rates, and has also broadened the distribution a bit. The small bumps at very high $\mathcal{R}$ are likely sampling artifacts, and should vanish if the MCMC is allowed to run for a much longer stretch of time in order to collect more samples.

The final results of our method are the posteriors on the actual mass distribution. First is the posterior on $p(m_1 \mid \alpha)$, shown in Figure 3.4 which is nearly identical to the mass distribution estimates used in [2]. Since this is a posterior on a distribution, we had to find a way to visualize a distribution of distributions. We computed $p(m_1 \mid \alpha)$ on a fine grid of $m_1$ values, for every value of $\alpha$ the MCMC drew from our posterior. Then for each $m_1$ value, we computed the median and $\pm 2\sigma$ credible regions for $p(m_1 \mid \alpha)$, which is shown in the figure as the dark line and the two shaded regions. Using the same approach, but now incorporating $\mathcal{R}$, we computed a posterior on $\rho(m_1 \mid \mathcal{R}, \alpha)$, shown in Figure 3.4. Since the difference between $p$ and $\rho$ is a multiplication by $\mathcal{R}$, one might expect it to simply scale the distribution up, but since we actually have a distribution on $\mathcal{R}$ this is not true. One might also expect the shape of the distribution to stay the same, except with the uncertainty on the distribution broadened, but the correlations between $\mathcal{R}$ and $\alpha$ actually allow the shape of the distribution to change to an extent. The distribution becomes somewhat less steep at high $m_1$, and the size of the credible regions become less variable as $m_1$ changes.

## 3.3 Simple extensions to the power law mass distribution

As was mentioned earlier this chapter, the choice of a power law for our mass distribution was primarily to remain consistent with existing LIGO publications. However, this is only a starting point. The next logical step is to take all of the parameters which were given assumed fixed values (i.e., $m_{\min}$ and $M_{\max}$), and instead make them free parameters. This will add more dimensions to our posterior distribution on $\Lambda$, which will make the results more honest, but also less constrained. It will also allow us to see if there is evidence for a mass gap either at low or high mass for black holes, as is discussed in [104, 62].

In addition to making the power law distribution more flexible, we can also investigate families of distributions with more flexibility. A logical next step would be to use polynomials.

A good way to handle this is to allow for unnecessarily high order polynomials, but then impose a smoothness prior which will result in low order polynomials being favored. As with the power law, we can make $m_{\min}$ and $M_{\max}$ parameters of the distribution, and investigate any mass gaps. Evidence of a mass gap under a polynomial model will be more believable, as it's quite likely that a power law is simply a bad model and biasing our results, whereas polynomials offer much more flexibility.

We are also ultimately interested not only in mass distributions, but spin distributions. In a future study we will extend our sample-spaces to include a spin dimension (measured by $\chi_{\text{eff}}$) and add more degrees of freedom to the distribution.

# Chapter 4

# Method III: Generic density estimators

One can try describing the binary black hole mass and spin distributions using every model they can think of. However, the number of possible distributions is uncountably infinite, with infinite degrees of freedom, making it a truly impossible task to perform perfectly. The only way one could perfectly model the full population is by observing every binary black hole merger in both the past and future of the entire Universe, with absolutely perfect measurements. Science fiction scenarios aside, this will never happen. However, finding the exact distribution is not important so long as we are close. Restricting ourselves to simple families of distributions like power laws won't get us very far, unless the Universe is unexpectedly simple and we are extremely lucky (i.e., power laws won't get us very far). Instead, we need to investigate more generic families of distributions, with many degrees of freedom.

The approach we will take is to use a mixture of simple parametric distributions. A mixture distribution can be thought of as a weighted average of multiple unique distributions, or even as a series expansion, where each of the basis functions is a probability distribution. In particular, we use a mixture of Gaussian distributions (also known as Gaussian mixture models or GMM's), which we describe in more detail in Section 4.1. Gaussian mixtures, as opposed to other mixtures, have also been studied and applied extensively, so there is a large body of literature for us to leverage. Of particular use is a method from the literature called "extreme deconvolution" [105], which allows one to make inferences about an underlying distribution when the observations have Gaussian measurement error. As we will explain further in Section 4.1, some of the key parameter estimates of LIGO sources have highly Gaussian measurement errors [106], making this a natural fit.

There are a number of alternatives one might consider over Gaussian mixtures, but they suffer shortcomings that make them undesirable, as detailed in Section 4.2.

In Sections 4.3 and 4.4, we will look at two examples of GMMs in action, by applying them to synthetic populations. In Section 4.3, the synthetic population is itself a GMM, whereas in Section 4.4, it is not, demonstrating the utility of GMM's as generic density estimators.

Finally, in Section 4.5, we will discuss some of the limitations of the current methodology, and improvements that we will make going forward.

## 4.1 Overview of Gaussian mixture models and their application to compact binaries

Consider some arbitrary parametric family of probability distributions, $f(x \mid \alpha)$, where $\alpha$ is its (possibly multidimensional) parameter. A $K$-component mixture of these distributions can be written as [107]

$$p(x \mid (w_1, \alpha_1), \ldots, (w_K, \alpha_K)) = \sum_{k=1}^{K} w_k f(x \mid \alpha_k), \tag{4.1}$$

where $w_1, \ldots, w_K$ are weights satisfying $w_k \geq 0$ and $\sum_k w_k = 1$. One can think of this as a basis function expansion, similar to a Fourier series, except instead of basis functions like $f(x \mid \alpha_k) = e^{ikx}$, our basis functions must be proper probability distributions, and our series coefficients must be non-negative and sum to unity. Solving for the parameters of the expansion need not be a linear problem, as we allow for $f(x \mid \alpha)$ to have arbitrary dependence on $\alpha$ – and in fact the case we're interested in is non-linear.

One can also think of each $f(x \mid \alpha_k)$ as describing the distribution of a sub-population, and the $w_k$ being the probability of a random draw from the full population being taken from the $k$th sub-population. So if we had a mixture distribution $p(x) = (1/2)f(x \mid \alpha_1) + (1/2)f(x \mid \alpha_2)$, then a random draw from $p$ would have a 50% chance of being drawn from the distribution $f(x \mid \alpha_1)$, and a 50% chance of being drawn from $f(x \mid \alpha_2)$.

We focus henceforth on mixtures of Gaussian distributions, i.e., we set $f(x \mid \alpha) = \mathcal{N}(x \mid \mu, \sigma^2)$ when $x$ is 1D, and $f(\boldsymbol{x} \mid \alpha) = \mathcal{N}(\boldsymbol{x} \mid \boldsymbol{\mu}, \boldsymbol{\Sigma})$ when $\boldsymbol{x}$ is multi-dimensional. For reference, a 1-dimensional Gaussian's probability density is given by [107]

$$\mathcal{N}(x \mid \mu, \sigma^2) = \frac{1}{\sqrt{2\pi\sigma^2}} \exp\left[-\frac{(x-\mu)^2}{2\sigma^2}\right], \tag{4.2}$$

where $\mu$ is its mean, and $\sigma^2$ is its variance. A $d$-dimensional Gaussian's probability density is given by [107]

$$\mathcal{N}(\boldsymbol{x} \mid \boldsymbol{\mu}, \boldsymbol{\Sigma}) = \frac{1}{\sqrt{(2\pi)^d \det \boldsymbol{\Sigma}}} \exp\left[-\frac{1}{2}(\boldsymbol{x}-\boldsymbol{\mu})^\top \boldsymbol{\Sigma}^{-1} (\boldsymbol{x}-\boldsymbol{\mu})\right], \tag{4.3}$$

where $\boldsymbol{\mu}$ is the mean vector, and $\boldsymbol{\Sigma}$ is the covariance matrix.

The general form for a $K$ component Gaussian mixture model, much like Equation 4.1, can be expressed as [107]

$$\mathrm{GMM}(\boldsymbol{x} \mid (w_1, \boldsymbol{\mu}_1, \boldsymbol{\Sigma}_1), \ldots, (w_K, \boldsymbol{\mu}_K, \boldsymbol{\Sigma}_K)) = \sum_{k=1}^{K} w_k \mathcal{N}(\boldsymbol{x} \mid \boldsymbol{\mu}, \boldsymbol{\Sigma}). \tag{4.4}$$

The free parameters in this expression are the $\boldsymbol{\mu}_k$'s (with $K \times d$ unique components), $\boldsymbol{\Sigma}_k$'s (with $K \times d(d+1)/2$ unique components), and $K-1$ of the $w_k$'s (as $\sum_k w_k = 1$ uniquely determines one of the components if the rest are known). Estimating a posterior distribution on all of these parameters is going to be very tricky, as it will be a very high dimensional posterior ($K[d(1+(d+1)/2)+1]-1$ dimensions to be precise). Furthermore, the posterior would have many modes. Since the precise labeling of the components is arbitrary, one can exchange any two of the components, and the resulting distribution will be unchanged. This means the posterior will have peaks for every possible relabeling. Also, there is the issue of choosing the number of Gaussians, $K$. To perform sampling via MCMC, one would need to construct a scheme where Gaussians may be created or destroyed at each step in the chain (e.g., through reversible-jump MCMC [108]), or otherwise run an MCMC for a range of $K$ values and reweigh the samples. To avoid these issues, at the cost of some rigor, we do not take a fully Bayesian approach here, in contrast to Chapter 3. Instead, we take a simple frequentist/maximum likelihood approach using the well established expectation maximization algorithm (EM algorithm) [109, 107].

LIGO does not perfectly measure the parameters of the sources it detects [55, 8, 9, 3]. This means that any attempt we make to infer the underlying population's distribution must somehow deconvolve this measurement error. One variant of the EM algorithm, called "extreme deconvolution" [105], addresses precisely this issue. The method is limited in its scope, however, to data whose measurement errors are Gaussian. Nature has been kind to us, as there exist coordinate systems in which the masses and spins of compact binaries are well approximated as Gaussians [106]. These coordinates are $(\mathcal{M}_\mathrm{c}, \eta, \chi_{1,z}, \chi_{2,z})$, where $\eta$ is the symmetric mass ratio, $\eta = m_1 m_2 / M^2$ ($M = m_1 + m_2$), $\mathcal{M}_\mathrm{c}$ is the chirp mass, $\mathcal{M}_\mathrm{c} = \eta^{3/5} M^{-1/5}$, and $\chi_{1,z}$, $\chi_{2,z}$ are the components of the two objects' dimensionless spins perpendicular to the orbital angular momentum. These coordinate systems are finite in extent (e.g., no negative masses), which can cause some issues, which we discuss in Section 4.5.

One thing not addressed by the EM algorithm, nor its variants, is the number of Gaussian components, $K$, to use in the mixture model. Using too few components will result in an overly smoothed out distribution, which misses features of the underlying distribution. Using too many components can cause several issues. First, if the number of components exceeds the number of data points, it becomes an over-determined problem, and the reconstructed distribution will likely be nonsensical. Even if the number of data points exceeds the number of mixture components, it is possible that the underlying distribution is simpler than the GMM being used to approximate it, likely resulting in a reconstructed distribution containing artificial features. To find the sweet spot for the number of mixture components, we estimate a GMM for a range of values of $K$, and for each of the resulting models, we compute the

Bayesian information criterion (BIC) [110]. The BIC can be written as [107]

$$\text{BIC} = (1/2)(\# \text{ of free parameters}) \log(N) - \log \hat{\mathcal{L}}, \tag{4.5}$$

where $\hat{\mathcal{L}}$ is the likelihood of the best fit model (i.e., the product of our GMM's probability density evaluated at each of the observed data points), $N$ is the number of data points, and the number of free parameters for a GMM is, as stated earlier, $K[d(1+(d+1)/2)+1]-1$. Note that a better fit will result in a higher $\log \hat{\mathcal{L}}$, reducing the BIC, and adding more parameters will increase the BIC. Therefore, the model which minimizes the BIC is the one which strikes a balance between a good fit and a simple model, so we use that model in our estimates.

## 4.2 Alternatives to Gaussian mixtures and their pitfalls

Mixtures of Gaussian distributions are far from the only generic density estimators, so one might ask why we did not choose an alternative to work with. In this section, we hope to convince you that they are well suited to this specific problem, and the alternatives have flaws that make GMM's the most attractive.

The density estimator with the widest recognition is the histogram. It is simple enough to be explained to young children, or computed on the back of an envelope: divide your data into bins, and count the fraction of data points that fall into each bin. This simplicity is very valuable when it comes to interpreting the results, so why don't we use this instead of GMM's?

First of all, the simple algorithm of dividing into bins and counting won't work here. LIGO does not perfectly measure the binary parameters; it has non-negligible measurement error. One could try modifying the algorithm: instead of simple counting, measure the total probability mass of each event's posterior contained within each bin. This would only produce the observed, noisy distribution, which is not particularly interesting. In order to get at the measurement error deconvolved distribution, one would likely have to take a fully Bayesian approach instead, which is going to have many degrees of freedom, and scale to higher dimensions (e.g., joint mass-spin distributions) poorly.

Furthermore, histograms inherently suffer a number of faults, most notably their sensitivity to the choice of binning [111]. Many textbook examples exist, including some in [111], which demonstrate qualitative changes in histograms simply by changing the bin width or shifting the bin centers slightly. For instance, peaks can appear and disappear.

Another method, which is often seen as a step up from a histogram, is the kernel density estimator (KDE). In this method, one places a small amount of probability density around each observed point, $x_n$, distributed according to a kernel function $\mathcal{K}$ [107]

$$p(x) \approx \frac{1}{N} \sum_{n=1}^{N} \mathcal{K}(x - x_n). \tag{4.6}$$

As in the case of the histogram, this only produces the observed distribution, and will need to be modified when dealing with measurement error, likely through a fully Bayesian approach.

Even without the measurement error issue, these tend to do poorly when the distribution requires varying levels of complexity in different regions of parameter space **(CITATION NEEDED)**. For instance, consider a number of samples drawn from a one dimensional distribution $p(x)$. If $p(x)$ is very smooth and featureless for large $x$, but bumpy and feature rich for small $x$, a KDE is not going to reconstruct $p(x)$ well until a very large number of samples are observed. The reason is that the kernel, which decides the shape of the probability added at each sample, is the same everywhere, and therefore if we need sharp, narrow contributions at small $x$ and broad contributions at large $x$, no kernel is going to work effectively everywhere. One will have to perform a coordinate transformation on the data first (perhaps in this scenario, $x \mapsto \log x$ would be effective) in order to make the estimator accurate. This of course requires some fore-knowledge about the distribution we are trying to estimate, which is sub-optimal.

Using a mixture of Gaussians is actually rather similar to a KDE in a sense, especially if one uses a Gaussian kernel. However, with GMM's, the mixture components have both their locations $\boldsymbol{\mu}$ and scaling $\boldsymbol{\Sigma}$ as free parameters, which addresses the issue with varying levels of complexity across the parameter space. In addition, instead of the data points determining the locations of the Gaussians, in a GMM it is in principle the Gaussians which determine the locations of the data points, although we ultimately have to solve the inverse problem.

## 4.3 Example I: Identifying sub-populations in simple models

We now take a look at a simple example, where we use samples from a GMM to reconstruct itself. However, we start with some additional motivation behind the particular choice of example.

One advantage of the GMM is that it acts not only as a density estimator, but also as a clustering algorithm. Consider a two component GMM:

$$p(x) = w_1 \mathcal{N}(x \mid \mu_1, \sigma_1^2) + w_2 \mathcal{N}(x \mid \mu_2, \sigma_2^2). \tag{4.7}$$

Given an observation from this distribution, $x^*$, one could compute the probability that it was drawn from the $k$th component (where $k \in \{1, 2\}$) by computing the "responsibility" [107]

$$\Pr[x^* \text{ is from component } k] = r_k(x^*) = \frac{w_k \mathcal{N}(x^* \mid \mu_k, \sigma_k^2)}{w_1 \mathcal{N}(x^* \mid \mu_1, \sigma_1^2) + w_2 \mathcal{N}(x^* \mid \mu_2, \sigma_2^2)}. \tag{4.8}$$

One could also come up with a binary scheme, associating $x^*$ with component 1 if $r_1(x^*) > 50\%$, and component 2 if $r_2(x^*) = 1 - r_1(x^*) > 50\%$. For two Gaussians which are highly non-overlapping (i.e., the regions bound by $\mu_1 \pm n\sigma_1$ and $\mu_2 \pm n\sigma_2$, for a large enough $n$ – e.g., $n = 5$ – are non-overlapping), this would be a very effective scheme. However, for highly similar Gaussians, it would be better to simply report the responsibilities.

In most real world scenarios, $p(x)$ will not be known. Instead, we will have a set of observations $\{x_1, \ldots, x_N\}$ which were drawn from $p(x)$. Also, it is highly unlikely that $p(x)$ is itself representable as a finite-component GMM. However, as was stated before, one can approximate $p(x)$ as a GMM given the samples $\{x_1, \ldots, x_N\}$ alone. If the GMM approximation to $p(x)$ has some overlapping components – as it likely will if $p(x)$ is not a non-overlapping GMM itself – then an additional clustering scheme can be applied to group nearby Gaussians together.

For this section, we look specifically at the simplest case, a synthetic population which is

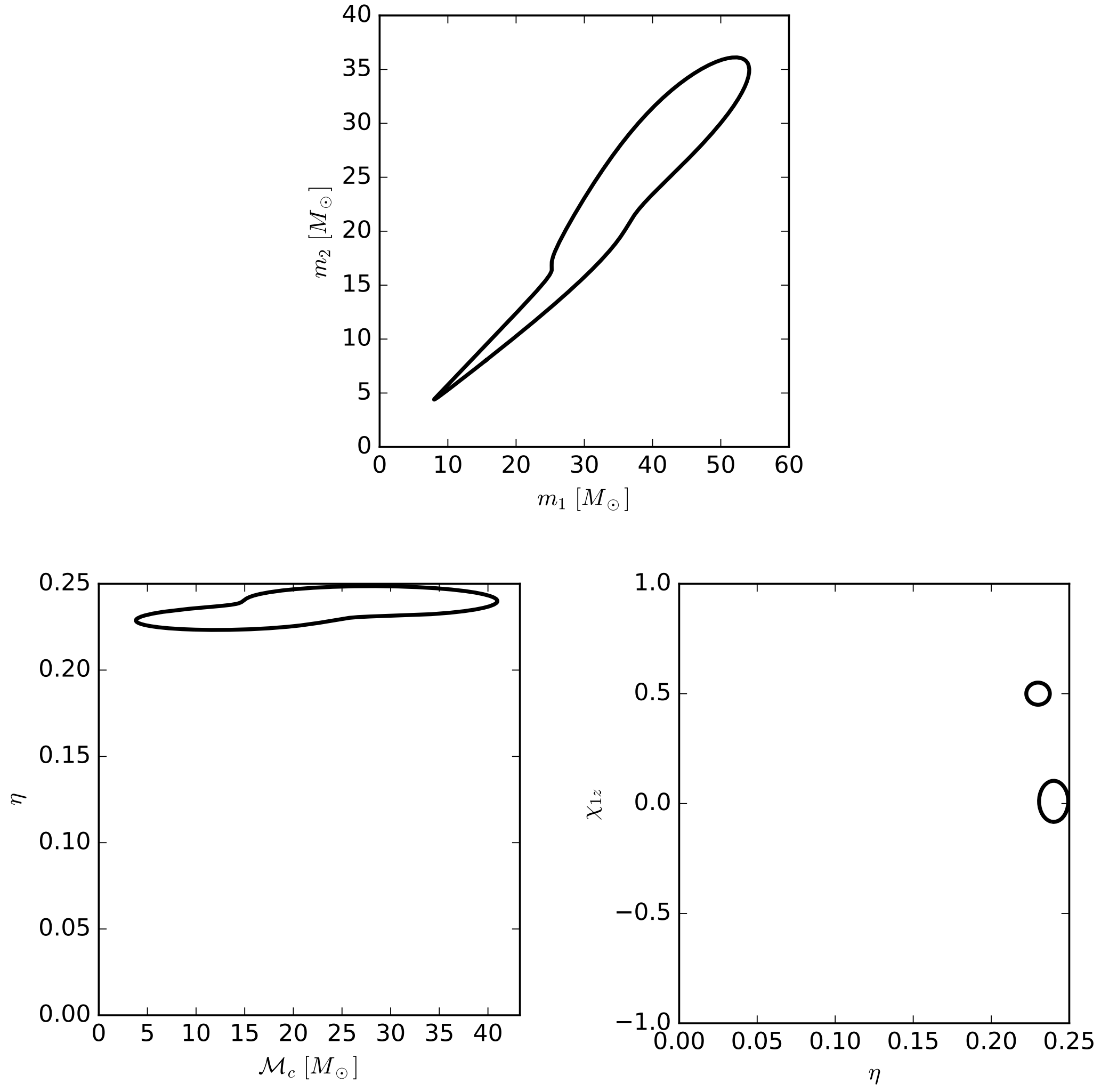

Figure 4.1: 90% confidence intervals for synthetic population, projected into several coordinate systems.

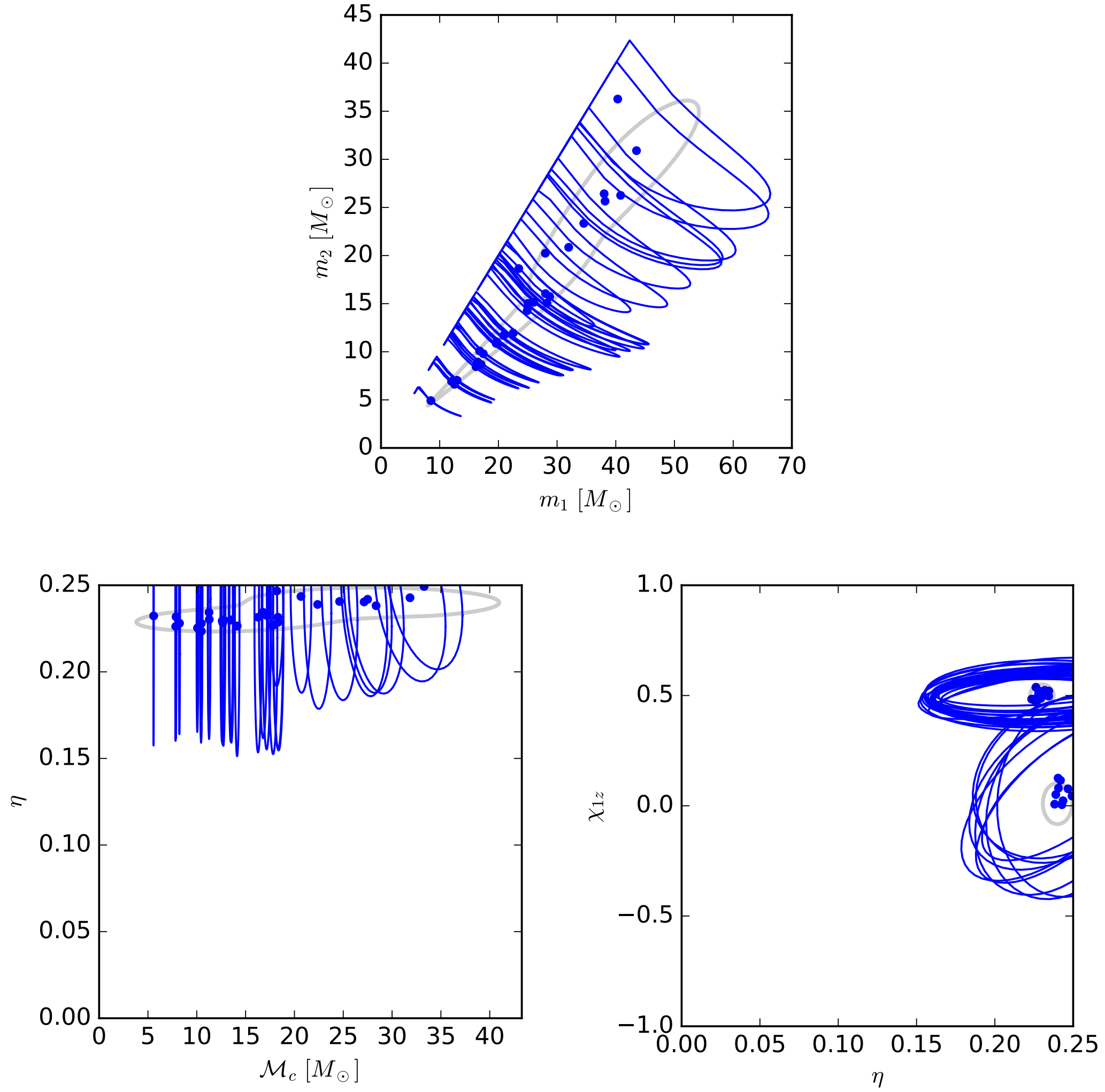

Figure 4.2: Synthetic population (black) with 30 random events drawn (blue). 90% confidence interval for each event's likelihood, approximated via Fisher matrix, are shown.

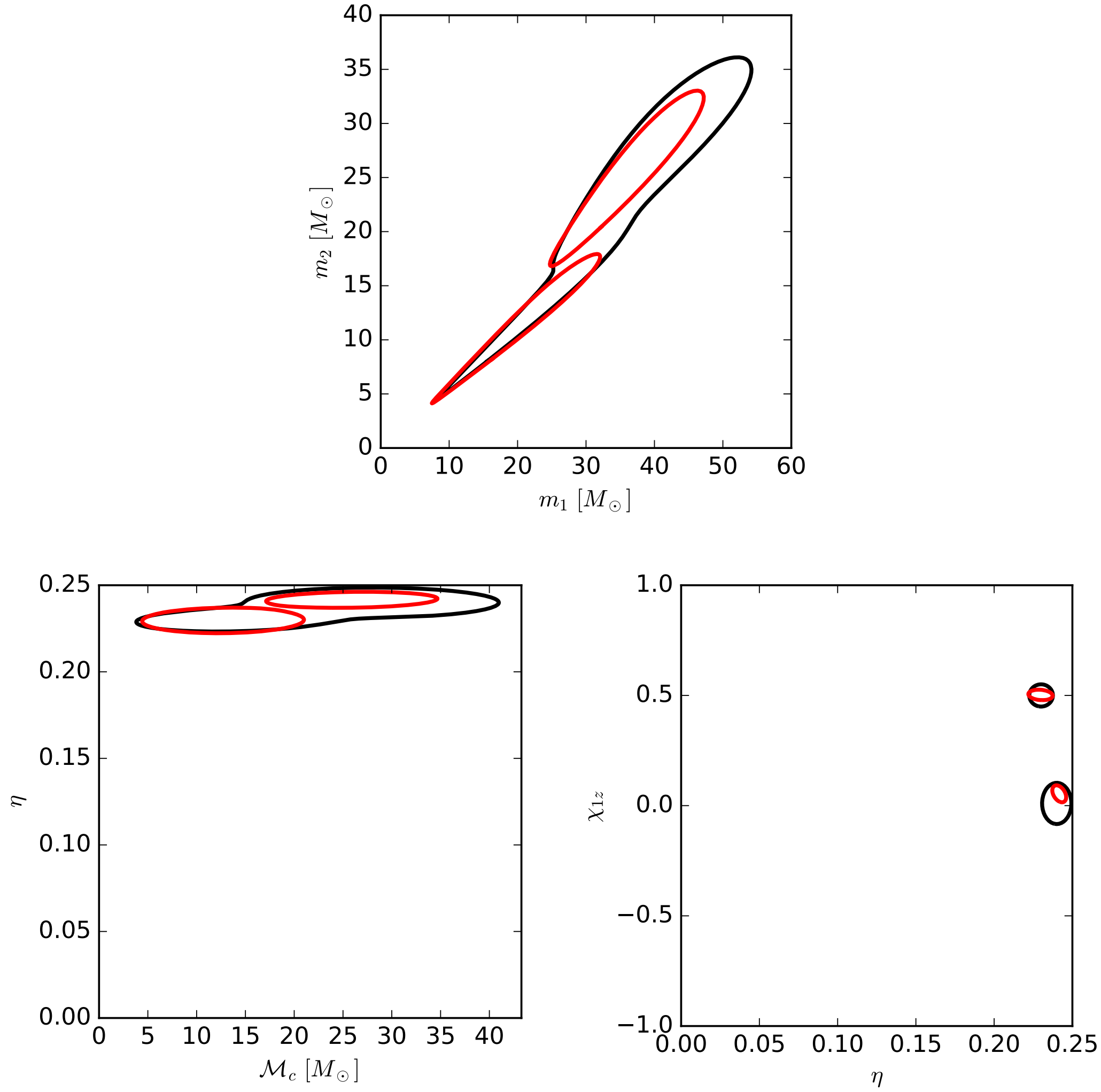

Figure 4.3: Synthetic population (black) and reconstructed population (red) shown with 90% confidence interval contours. Reconstruction was made from the 30 random events shown in 4.2.

| Component | Weight | Mean $\boldsymbol{\mu}$ | | Covariance $\boldsymbol{\Sigma}$ | |
|---|---|---|---|---|---|
| | | Variable | Value | Variables | Value |
| 1 | 50% | $\mathcal{M}_{\rm c}$ | $28\,{\rm M}_{\odot}$ | $(\mathcal{M}_{\rm c}, \mathcal{M}_{\rm c})$ | $(7.5\,{\rm M}_{\odot})^2$ |
| | | $\eta$ | 0.24 | $(\eta, \eta)$ | $0.005^2$ |
| | | $\chi_{1,z}$ | 0.01 | $(\chi_{1,z}, \chi_{1,z})$ | $0.05^2$ |
| 2 | 50% | $\mathcal{M}_{\rm c}$ | $13.9\,{\rm M}_{\odot}$ | $(\mathcal{M}_{\rm c}, \mathcal{M}_{\rm c})$ | $(4.5\,{\rm M}_{\odot})^2$ |
| | | $\eta$ | 0.23 | $(\eta, \eta)$ | $0.003^2$ |
| | | $\chi_{1,z}$ | 0.5 | $(\chi_{1,z}, \chi_{1,z})$ | $0.02^2$ |
| | | | | $(\mathcal{M}_{\rm c}, \eta)$ | $0.05^2{\rm M}_{\odot}$ |

Table 4.1: Parameters of two sub-population synthetic model used in this section. All covariance terms not listed are either implicit due to the matrix's symmetry (i.e., $\Sigma_{ij} = \Sigma_{ji}$) or are zero. The Gaussians are truncated according to the limits of ($\mathcal{M}_{\rm c}$,$\eta$,$\chi$), and the components are re-weighted such that the truncation does not alter the total probability mass allocated to each component.

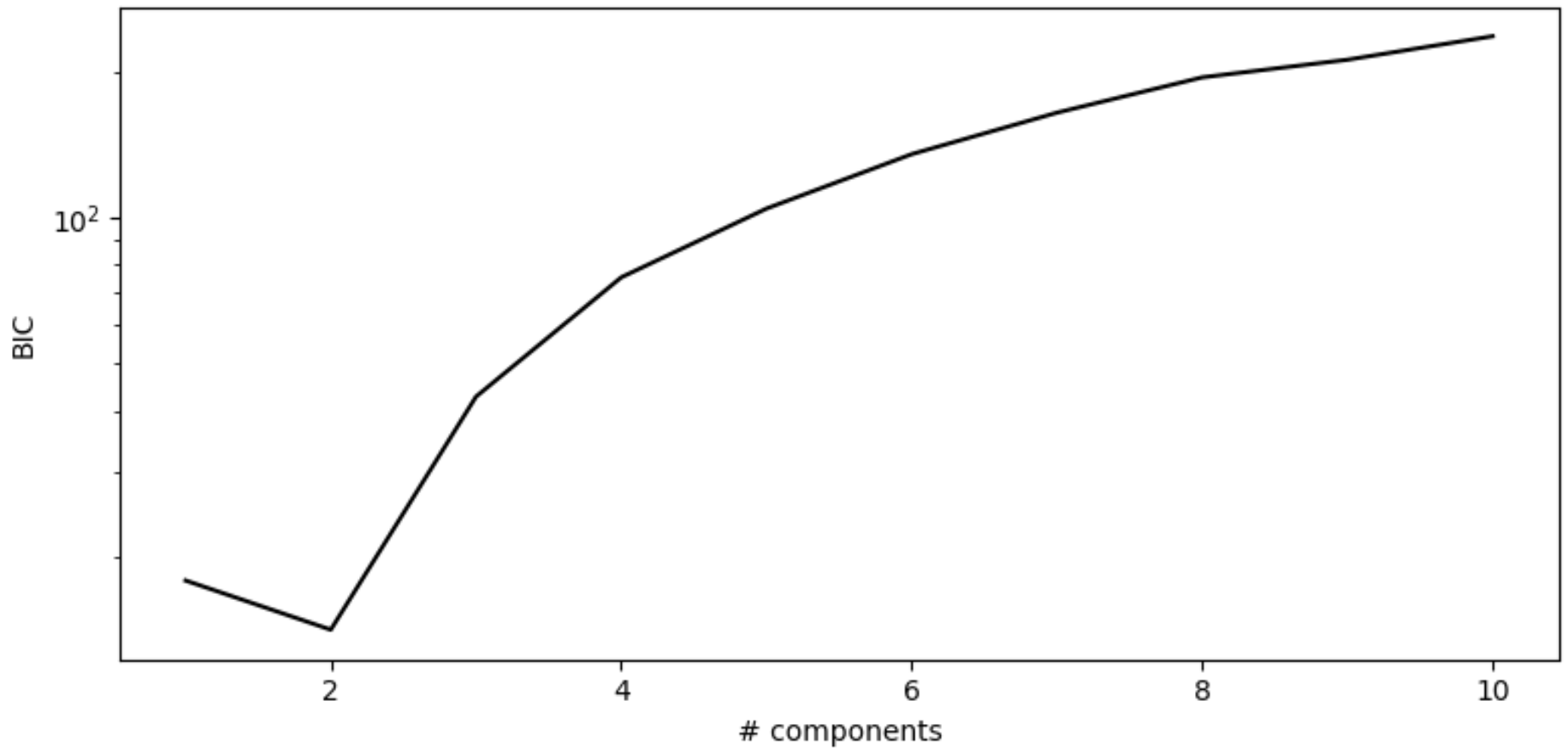

Figure 4.4: Bayesian information criterion as a function of the number of Gaussians used in reconstructing the two component model from 30 samples. It is minimized for the correct number of components: two.

*precisely* a (truncated) two component GMM in ($\mathcal{M}_{\rm c}$, $\eta$, $\chi_{1,z}$) coordinates. The parameters of this mixture model are shown in Table 4.1, and the 90% probability contours in several 2D projections are shown in Figure 4.1. One of the sub-populations has high mass and low spin, whereas the other has low mass and high spin, motivated very loosely by GW150914 [4] and GW151226 [8], respectively. The specific values were tweaked to ensure most of the probability was not outside the range of allowed values of ($\mathcal{M}_{\rm c}$,$\eta$,$\chi$). We also designed the population to be overlapping in the $\mathcal{M}_{\rm c}$–$\eta$ plane, but separated in the $\mathcal{M}_{\rm c}$–$\chi_{1,z}$ and $\eta$–$\chi_{1,z}$ planes, so that spin information would allow us to distinguish the two sub-populations.

With this population model specified, we drew 30 random samples from the distribution, and gave them realistic measurement errors via the Fisher matrix technique [106]. The samples and their error ellipsoids can be seen in Figure 4.2.

We then estimated the underlying Gaussian parameters from the observed samples, using the extreme deconvolution method [105]. We repeated this with 1, 2, ..., 10 Gaussians, and evaluated the Bayesian information criterion (Equation 4.5) for each of these models, as shown in Figure 4.4. The BIC is smallest for the model with two components, which is the correct number. This reconstructed distribution is plotted alongside the true distribution in Figure 4.3. Despite the low number of samples, the reconstruction has successfully latched onto the two sub-populations' locations. However, it does not do a particularly good job reproducing the shapes of the sub-populations. This is not too surprising, considering that the model has 19 free parameters (1 for the weights, $2 \times 3$ for the means, and $2 \times 6$ for the covariance matrices) and we only have 30 noisy data points with which to make our inference. Clearly we will need many detections before this method will produce reliable reconstructions of compact binary parameter distributions.

Also, in this example we only looked at the observed distribution. Ideally we should start with an intrinsic distribution, selection bias our samples, and then reconstruct the intrinsic distribution from those biased samples. This will be done in future work, and is discussed in Section 4.5.

## 4.4 Example II: Reconstructing a power law with a mass gap

Now that we've seen how Gaussian mixtures can be used to reconstruct distributions which are themselves Gaussian mixtures, we will take a look at using them to reconstruct a distribution which is expressly not. For this we go back to the familiar case of a power law mass distribution, as seen in Section 3.2, with a few differences. First of all, this model is truly a power law in $m_1$ (i.e., $p(m_1) \propto m_1^{-\alpha}$), with no high mass turnoff due to constraints on $m_1 + m_2$, as in Equation 3.8. Second, we extended the mass range to $m_{\mathrm{min}} = 1\,\mathrm{M}_{\odot}$ and $m_{\mathrm{max}} = 100\,\mathrm{M}_{\odot}$, and added a mass gap between 3–4 $\mathrm{M}_{\odot}$. This is to test our ability to identify the mass gap between neutron stars and black holes that is believed to exist [104]. For the purposes of this test, we set the power law index to $\alpha = 1.05$.

We used this synthetic population to generate sample observations, with no measurement error, and for each set of samples we estimated the underlying distribution via a Gaussian mixture model. We varied the sample size from tens up to thousands, reconstructing the distribution from each sample, and repeating each case 500 times to quantify the variance of our maximum likelihood estimator. The distribution of results is shown for 30, 50, 100, and 1000 samples in Figure 4.5.

As the number of samples increases, the variance of our estimator is consistently reduced. However, it does not converge to the underlying distribution. For one, edge effects in our reconstruction method at both the high and low mass cutoffs cause us to underestimate the mass distribution at those endpoints. Similarly, around the mass gap we under-predict the mass distribution, and inside of the mass gap we over-predict the mass distribution. While it should drop to absolute zero inside of the gap, we instead see a mere downward bump in the distribution. This is due to the fact that Gaussian distributions are continuous and non-zero everywhere, which has the effect of smoothing out the mass gap. This is reminiscent of the Gibbs phenomenon, which was discovered in the context of approximating functions with discontinuous jumps using finite-order Fourier series [112, 113], and later generalized to other series expansions [114]. Since a GMM can be thought of as a smooth series expansion, and the mass gap is a discontinuity, the scenarios are similar.

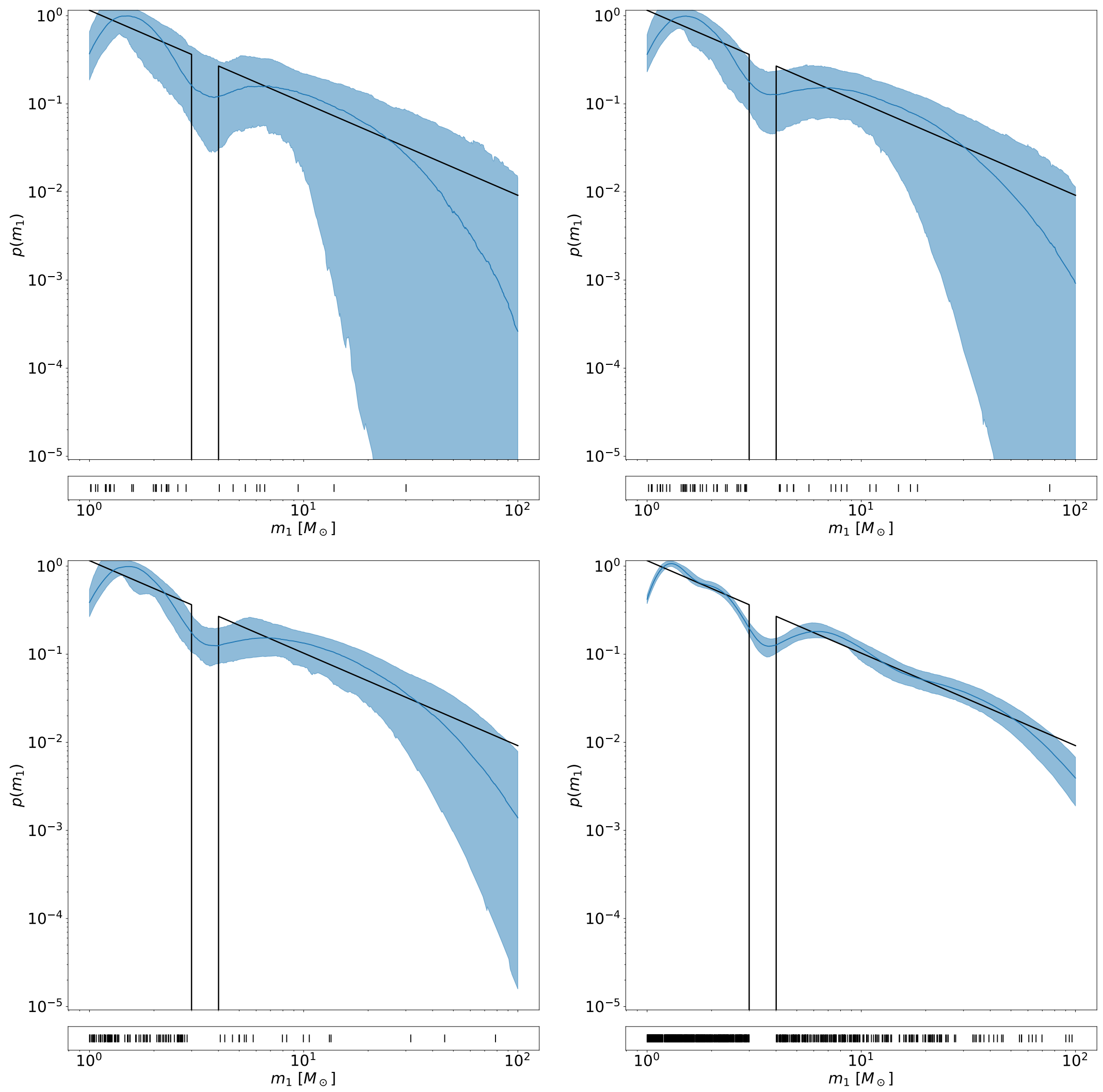

Figure 4.5: Underlying power law mass distribution (black) with mass gap between 3–4 $M_\odot$. Distribution reconstructed from 30 (upper left), 50 (upper right), 100 (lower left), and 1000 (lower right) samples. For each number of samples, 500 independent trials were run, with the samples from one of those trials shown below each plot. From these trials, the median and 90% confidence interval on our maximum likelihood estimate are shown as a solid line and shaded region, respectively. This shaded region is not to be confused with the posterior distribution – which was not computed – and does not account for our uncertainties, but instead quantifies the variance of our maximum likelihood estimator.

As was an issue in Section 4.3, we only reconstructed the observed population here. It is the intrinsic population that existing studies (e.g., [9]) have approximated as a power law, not the observed one. This is something that will be addressed in future work, as discussed in Section 4.5.

## 4.5 Pitfalls and improvements to GMM methodology

There are a few limitations to our approach, which will be better addressed in future work. Most importantly is the incorporation of selection effects (e.g., as discussed in Section A.2), for which we have some preliminary results, but which were not mature enough to include in this thesis. This can be incorporated into the EM algorithm by re-weighting the Gaussian components by the inverse of the selection bias.

Another issue is that our coordinate system has hard edges ($\mathcal{M}_\mathrm{c} > 0\,\mathrm{M}_\odot$, $0 < \eta \leq 1/4$, $-1 \leq \chi \leq +1$), whereas Gaussians (and therefore Gaussian mixtures) have non-zero probability until $\pm\infty$. The limits on $\mathcal{M}_\mathrm{c}$ only cause an issue for very low mass objects, which does not affect the scenarios we investigate in this thesis. The upper and lower limits on $\eta$ are only issues for maximally spinning binaries, which we also do not investigate in this thesis. The lower limit on $\eta$ is an issue for binaries with very dissimilar component masses, which we also avoid. However, the high limit ($\eta \leq 1/4$) is very much an issue. To deal with this, we have all of our inferences initially assume $\eta$ extends to $\pm\infty$, and after a Gaussian mixture model has been inferred from the data, we re-weigh each component such that its total probability mass is preserved. A component with weight $w_k$ should have total probability mass $w_k$, as $\int_{-\infty}^{\infty} w_k \, \mathcal{N}(\eta \mid \mu_k, \sigma_k^2) \mathrm{d}\eta = w_k$. Applying the $\eta \leq 1/4$ cutoff (and ignoring the lower cutoff) changes this to (using [107] for the Gaussian integral)

$$\int_{-\infty}^{1/4} w_k \, \mathcal{N}(\eta \mid \mu_k, \sigma_k^2) \mathrm{d}\eta = w_k \, \left\{ \frac{1}{2} \left[ 1 + \mathrm{erf}\left( \frac{1/4 - \mu_k}{\sqrt{2\sigma_k^2}} \right) \right] \right\}, \tag{4.9}$$

where $\mathrm{erf}(x)$ is the error function. To change the total probability mass to $w_k$, we divide each Gaussian's weight by the term in curly braces in Equation 4.9.

This only corrects the mixture weights, but not the $\boldsymbol{\mu}_k$'s and $\boldsymbol{\Sigma}_k$'s. To properly address this, we should use a variant of the EM algorithm derived from first principles with truncated Gaussians. Some literature exists on this subject [115], and we will address this in future work.

# Chapter 5

# Discussion

In this thesis, we have developed and tested a number of methods to understand the population of binary black holes observed by LIGO. In this early stage of 3.9 detections, the constraints we can make are still weak. However, we have laid some of the ground work for big discoveries in the years to come, as LIGO and Virgo detect more black hole mergers, filling in the gaps in our understanding of the binary black hole population. If the upgrades planned for Advanced LIGO's third observing run are as effective as planned, we might reach the 100th detection in the next few years, at which point these methods will begin to see their full potential.

We can also extend these methods to other GW sources that LIGO has the potential to detect, namely the merger of two neutron stars and a neutron star–black hole binary. The physical models used in Chapter 2 already produce both types of sources, but we left them out of our analysis. Incorporating them into the analysis only requires some minor changes such as mass cutoffs. The specific parametric models used in Chapter 3 will need to be modified to account for things like a neutron star–black hole mass gap [104]. Simple power laws will not do well with neutron star populations, as we already know from electromagnetic surveys that this is a bad approximation **(NEED CITATION)**. More flexible distributions like the polynomials with smoothness priors discussed in Section 3.3 will likely do a better job. The methods from in Chapter 4 are still not fully ready for even binary black hole populations, so some improvements are still needed there. In order to deal with a potential mass gap, we will need to overcome the challenges seen in Section 4.4. Otherwise, we can focus on a single source morphology, and infer just the single distribution (e.g., mass distribution for binary black holes done separately from binary neutron stars), although this will break down once a single ambiguous event is detected (e.g., mass parameter estimates span both expected ranges for black holes and neutron stars).

Even with the current 3.9 detections, the results from Chapters 2 and 3 give us some new insights. In Chapter 2, our most significant result is that, if we assume two types of black holes exist, those with high natal spins and those with low birth spins (explained in more detail in Appendix B.2), then the expected fraction of each strongly favors the low spin case, as shown in the top panel of Figure 2.7. The degree to which it is favored decreases as we increase the strength of supernova kicks, $\sigma_{\mathrm{kick}}$, but it is still favored in every case we investigated. We also show that modest supernova kicks are favored, given the 3.9 events, although that is subject to change with more detections.

In Chapter 3, we have made some advancements over existing work [9, 3] on the power law mass distribution. It was already known that fixing the power law index $\alpha$ when doing rate estimation would bias the results, and now we have the first results without that bias. Of course, there are still other biases, such as that created by assuming a power law to begin with, or by fixing the upper and lower limits on black hole masses, which will be alleviated when we make the improvements mentioned in Section 3.3.

By the time the third Advanced LIGO observing run begins (circa end of 2018 / beginning of 2019 [116] (**NOTE: citation is for latest published version (2016), but referencing latest version (September 2017)**)), the methods described in this thesis will have reached maturity, and be prepared for the anticipated high detection rate of GW events. Only time will tell what we find.

# Appendices

# Appendix A

# Miscellaneous formulae

This appendix describes a number of formulae which are used throughout the text, but which are neither critical enough to be described in the main text, nor long enough to be given their own appendices.

## A.1 Basics of Bayesian inference

At the heart of Bayesian inference lies one simple formula: Bayes' theorem.

$$p(\mathcal{M} \mid \mathcal{D}) = \frac{p(\mathcal{D} \mid \mathcal{M})\, p(\mathcal{M})}{p(\mathcal{D})} \tag{A.1}$$

While this expression holds true for any $\mathcal{M}$ and $\mathcal{D}$, in Bayesian inference one has $\mathcal{M}$ denote a model and $\mathcal{D}$ denote some data. The end goal of a Bayesian inference problem is to determine the probability of various models given the observed data, and in that sense one can think of $\mathcal{D}$ as a constant, and $p(\mathcal{M} \mid \mathcal{D})$ as a function of one's choice of model, $\mathcal{M}$. This function is often referred to as the "posterior" distribution, as it describes our degree of belief in each model after observing the data (*a posteriori*). This is in contrast to the probability of each model *prior* to observing any data, $p(\mathcal{M})$. This is called the "prior" distribution, and is often distinguished by writing it as $\pi(\mathcal{M}) \equiv p(\mathcal{M})$. Similar to the prior is the "evidence", $p(\mathcal{D})$, which describes the probability of observing the data under all possible models. However, since our data are not changing (and if they did change, we would simply redo our analysis), $p(\mathcal{D})$ is constant, and we simply absorb it into a proportionality constant. Last but not least is the probability of our data being that which is observed, given that a particular model is correct, $p(\mathcal{D} \mid \mathcal{M})$. This is referred to as the "likelihood", and is often distinguished by writing it as $\mathcal{L}(\mathcal{M}) = p(\mathcal{D} \mid \mathcal{M})$ (sometimes the $\mathcal{L}$ is written differently, e.g. as $\ell$, when multiple likelihoods are involved in a single calculation, and need to be distinguished typographically). This can be thought of as our forward model – given a particular base truth, what do we expect the distribution of possible realizations of our data to be? Now evaluate that distribution at the data that were actually observed and you have $\mathcal{L}(\mathcal{M})$.

Putting all of this together, we can write Bayes' theorem in a different way, which matches its use in practice.

$$p(\mathcal{M} \mid \mathcal{D}) \propto \mathcal{L}(\mathcal{M}) \, \pi(\mathcal{M}) \tag{A.2}$$

Since this expression is only a proportional relation, it also adds the convenience of only needing to specify $\mathcal{L}$ and $\pi$ up to a proportionality constant, taking care that it is truly constant (i.e., is not a function of $\mathcal{M}$). This turns out to be quite convenient.

Now, Equation A.2 would only be useful in the simplest of problems if not for the workhorse of Bayesian inference – Markov chain Monte Carlo (MCMC). This is a general purpose algorithm for sampling from a distribution, when the distribution is only known up to a proportionality constant. It accomplishes this by only ever using ratios of probabilities, e.g., $p(\mathcal{M} \mid \mathcal{D}) / p(\mathcal{M}' \mid \mathcal{D})$. Throughout this thesis, we make use of a particular MCMC variant, the Affine Invariant MCMC Ensemble sampler [117], and its Python implementation `emcee` [118]. While this does not give us the posterior itself – only samples from it – we can still estimate the posterior from the samples through density estimation, or use the samples in place of the posterior by doing Monte Carlo calculations.

## A.2 Detection volumes

When dealing with detection rates in the context of gravitational wave observations, effective detection volumes often come up. As a gravitational wave detector's sensitivity improves, it can detect ever more distant sources. The maximum distance at which a source can be detected is called the horizon distance. This is a function not only of the detector's efficiency, but also the intrinsic and extrinsic properties of the source. For instance, a more massive binary can be detected at a greater distance than a less massive one, as the gravitational wave amplitude is increased. Another factor is the binary's orientation – a face-on binary will produce a greater amplitude wave than an edge-on binary.

We will quantify all of this through the effective observed volume, $\langle V\rangle(\lambda)$, where $\lambda$ represents all of the intrinsic binary parameters. The $\lambda$ dependence of $\langle V\rangle$ comes entirely from the detection probability, $p_{\mathrm{det}}(z,\lambda)$. This is the probability of detecting a source located at a redshift of $z$ with intrinsic parameters $\lambda$, assuming an isotropic distribution of sources on the sky, with isotropic orientations. What one means by detection can take different levels of rigor here. Detection should mean that the search pipelines identify the source, and identify it as astrophysical in origin. However, for the purposes of a simple calculation, we take this to mean that the expected signal-to-noise ratio $\rho$ exceeds some threshold $\rho_{\mathrm{th}}$ (we choose $\rho_{\mathrm{th}} = 8$ as is common in the literature [33]). The expected SNR can be computed via [119]

$$\rho^2 = 4\int_0^\infty \frac{|\tilde{h}(f)|^2}{S_n(f)}\mathrm{d}f, \tag{A.3}$$

where $\tilde{h}(f)$ is the Fourier transform of the strain from the signal, and $S_n(f)$ is the noise power spectral density of the detector. For a multiple detector ($N_{\mathrm{det}}$) network, the total SNR can be computed by adding the individual SNR's in quadrature

$$\rho^2 = \sum_{i=1}^{N_{\mathrm{det}}} \rho_i^2. \tag{A.4}$$

With this in hand, the effective detection volume $\langle V \rangle$ can be computed as [33]

$$\langle V \rangle(\lambda) = \int_0^\infty \frac{1}{1+z} \, p_{\mathrm{det}}(z, \lambda) \, \frac{\mathrm{d}V_c}{\mathrm{d}z} \, \mathrm{d}z. \tag{A.5}$$

This can, e.g., be computed on a grid of $\lambda$ values and then interpolated. Each value of $(\lambda, z)$ is computationally intensive, as it requires computing $p_{\mathrm{det}}$, which means generating synthetic waveforms for an isotropic distribution of $\lambda$'s, and counting the fraction at that $z$ which result in $\rho > \rho_{mathrmth}$.

Finally, not only do we care about the observed volume, but also the amount of time spent observing that volume, $T$. Multiplying the two gives the sensitive space and time volume $\langle VT \rangle(\lambda) = \langle V \rangle(\lambda) \cdot T$. With all of this in hand, the merger rate density, $\mathrm{d}N/\mathrm{d}V\mathrm{d}t\mathrm{d}\lambda$, can be converted into an expected observed count by multiplying by $\langle VT \rangle$ and integrating over the $\lambda$'s of interest

$$\mu = \int \langle VT \rangle(\lambda) \, \frac{\mathrm{d}N}{\mathrm{d}V\mathrm{d}t\mathrm{d}\lambda} \mathrm{d}\lambda. \tag{A.6}$$

Throughout this thesis, we took $\lambda = (m_1, m_2)$, and assumed zero spin. We also used a noise PSD from the time around GW150914. This is consistent with what was done for the rate calculations in the catalog paper for LIGO's first observing run [2].

# Appendix B

# Mixture of physical models

This appendix combines the multiple appendices in the paper which became Chapter 2.

## B.1 Approximating parameter distributions from finite samples

Our population synthesis techniques allow us to generate an arbitrarily high number of distinct binary evolutions from each formation scenario, henceforth indexed by $\Lambda$. Instead of generating individual binary evolution histories, we weigh each one by an occurrence rate, allowing it to represent multiple binaries. For our calculations, however, we instead require the relative probability of different binaries, not just samples from the distribution. We estimate this distribution from the large but finite sample of binaries available in each synthetic universe. We do not simply use an occurence rate-weighted histogram of all the samples. Histograms work reliably for any single parameter (e.g., $p(m_1|\Lambda)$), where many samples are available per potential histogram bin, but for high-dimensional joint distributions (e.g., $p(m_1, m_2, \theta_1, \theta_2, \chi_{\mathrm{eff}}|\Lambda)$), many histogram bins will be empty simply due to the curse of dimensionality.

In all our calculations, we instead approximate the density as a mixture of Gaussians, labeled $k = 1, 2, \ldots, K$, with means and covariances ($\boldsymbol{\mu}_k$, $\boldsymbol{\Sigma}_k$) to be estimated, along with weighting coefficients $w_k$, which must sum to unity. The density can therefore be written as

$$p(\boldsymbol{x}) \approx \sum_{k=1}^{K} w_k \, \mathcal{N}(\boldsymbol{x}|\boldsymbol{\mu}_k, \boldsymbol{\Sigma}_k), \tag{B.1}$$

where $\mathcal{N}(\cdot)$ represents the (multivariate) Gaussian distribution. We select the number of Gaussians $K$ by using the Bayesian information criterion.

To estimate the means and covariances of our mixture of Gaussians, we used the expectation maximization algorithm [120]; see, e.g., [107] for a pedagogical introduction. Specifically, we used a small modification to an implementation in `scikit-learn` [121], to allow for weighted samples in the update equation (e.g., adding weights to Eq. (11.27) in [107]).

Ideally we would simply approximate each formation scenario $\Lambda$'s intrinsic predictions $p(\mathbf{x}|\Lambda)$ with a mixture of Gaussians, using the merger rate for each sample binary as its weighting factor. However, all astrophysical indications suggest that more massive progenitors form more rarely, implying this procedure would result in a distribution that is strongly skewed in favor of the much more intrinsically frequent low mass systems; our fitting algorithm might end up effectively neglecting the samples with small weights. This would risk losing information about the most observationally pertinent samples, which due to LIGO's mass-dependent sensitivity are concentrated at the highest observationally accessible masses. Alternatively, for every choice of detection network, we can approximate each formation scenario's predictions *for that network.* If $TV(\mathbf{x})$ is the average sensitive 4-volume for the network, according to this procedure we approximate $V(\mathbf{x})p(\mathbf{x})$ by a Gaussian mixture, then divide by $V(\mathbf{x})$ to estimate $p(\mathbf{x})$. To minimize duplication of effort involved in regenerating our approximation for each detector network, we instead adopt a *fiducial* (approximate) network sensitivity model $V_{\rm ref}(\mathbf{x})$ for the purposes of density estimation. We adopt the simplest (albeit ad-hoc) network sensitivity model: the functional form for $V(\mathbf{x})$ that arises by using a single detector network and ignoring cosmology (i.e., $EV \propto \mathcal{M}_c^{15/6}$) [86]. The overall, nominally network- and run-dependent normalization constant in this ad-hoc model $V_{\rm ref}$ scales out of all final results.

## B.2 Hierarchical comparisons of observations with data

As described in Section 2.3.2, the population of binary mergers accessible to our light cone can be described as an inhomogeneous Poisson process, characterized by a probability density $e^{-\mu}\prod_k \mathcal{R}p(x_k)$ where $x_k = x_1 \ldots x_N$ are the distinct binaries in our observationally accessible parameter volume $\mathcal{V}$. In this expression, the expected number of events and parameter distribution are related by $\mu = \int dx\sqrt{g}\mathcal{R}p(x)$; the multidimensional integral $\int dx\sqrt{g}$ is shorthand for a suitable integration over a manifold with metric; and the probability density $p(x)$ is expressed relative to the fiducial (metric) volume element, but normalized on a larger volume than $\mathcal{V}$. Accounting for data selection [87], the likelihood of all of our observations is therefore given by Eq. (2.2).

To insure we fully capture the effects of precessing spins, we work not with the full likelihood – a difficult function to approximate in 8 dimensions – but instead with a fiducial posterior distribution $p_{\rm post} = Z^{-1}p(d_k|x)p_{\rm ref}(x_k)$, as would be provided by a Bayesian calculation using a reference prior $p_{\rm ref}(x_k)$. Rewriting the integrals $\int dx_k p(d_k|x)p(x_k|\Lambda)$ appearing in Eq. (2.2) using the reference prior we find integrals appearing in this expression can be calculated by Monte Carlo, using some sampling distribution $p_{s,k}(x_k)$ for each event (see, e.g., [122]):

$$\int dx_k p(d_k|x_k)p(x_k|\Lambda) = \frac{1}{N_k}\sum_s \frac{[p(d_k|x_k)p_{\rm ref}(x_{k,s})]p(x_{k,s}|\Lambda)}{p_{s,k}(x_{k,s})p_{\rm ref}(x_{k,s})}, \tag{B.2}$$

where $s = 1, \ldots N_k$ indexes the Monte Carlo samples used. One way to evaluate this integral is to adopt a sampling distribution $p_{s,k}$ equal to the posterior distribution evaluated using the reference prior, and thus proportional to $p(d_k|x_k)p_{\rm ref}(x_k|\Lambda)$. If for this event $k$ we have samples $x_{k,s}$ from the posterior distribution – for example, as provided by a Bayesian Markov chain Monte Carlo code – the integrals appearing in Eq. (2.2) can be estimated by

$$\int dx_k p(d_k|x_k)p(x_k|\Lambda) \simeq \frac{Z}{N_k}\sum_s \frac{p(x_{k,s}|\Lambda)}{p_{\rm ref}(x_{k,s})}, \tag{B.3}$$

We use this expression to evaluate the necessary marginal likelihoods, for any proposed observed population $p(x|\Lambda)$.

In the expression above, we need only consider *some* of the degrees of freedom in the problem. Notably, the probability distributions for extrinsic parameters like the source orientation, sky location, and distance will always be in common between our models and our reference prior. So will any Jacobians associated with changes of coordinate. Moreover, these assump-tions are independent of one another and of the intrinsic parameter distributions. Therefore, the ratio of probability densities $p(x|\Lambda)/p_{\rm ref}(x)$ usually has product form, cancelling term by term. We therefore truncate the ratio to only account for *some* of the degrees of freedom.

To verify and better understand our results, we can also *approximate* the likelihood function, using suitable summary statistics. As an example, **(author?)** [90] reproduce parameter estimates of GW150914 using a Gaussian approximation to the likelihood and the assumption of perfect spin-orbit alignment. Using this approximation, and a similar approximation for GW151226, we can alternatively approximate each integral appearing in the likelihood by using the (weighted) binary evolution samples $x_{k,A}$ and their weights $w_A$:

$$\int dx_k p(d_k|x_k)p(x_k|\Lambda) \simeq \frac{\sum_A w_A \hat{p}(d_k|x_{k,A})}{\sum_A w_A} \tag{B.4}$$

where $\hat{p}$ refers to our approximate likelihood for the $k$th event. Even though these likelihood approximations neglect degrees of freedom associated with spin precession, we can reproduce the observed mass and $\chi_{\rm eff}$ distributions reported in **(author?)** [90]. We used this approximate likelihood approach to validate and test our procedure. We also use this approach to incorporate information about GW170104, which was not available at the same level of detail as the other events.

As an example, we describe how to evaluate this integral in the case where $p(x_k|\lambda)$ is a mixture model $p(x|\lambda) = \sum_\alpha \lambda_\alpha p_\alpha(x)$, for $\lambda$ an array of parameters. In this case, all the integrals can be carried out via

$$\prod_k \int dx_k p(d_k|x_k)p(x_k|\lambda) = \prod_k \left[\sum_\alpha \lambda_\alpha \int dx_k p(d_k|x_k)p_\alpha(x_k)\right] = \prod_k \sum_\alpha \lambda_\alpha c_{\alpha,k} \tag{B.5}$$

where $c_{\alpha,k}$ are integrals we can compute once and for all for each event, using for example the posterior samples from some fiducial analysis. As a result, the observation-dependent factor in likelihood for a mixture model always reduces to a homogeneous $N$th-degree polynomial in the mixture parameters $\lambda_\alpha$. Bayes theorem can be applied to $\lambda$ to infer the distribution over mixture parameters. Depending on the mixture used, this calculation could incorporate a physically-motivated prior on $\lambda$.

We use a mixture model approach to hierarchically constrain the spin magnitude distribution implied by our data. In our approach, we first consider models where both spin magnitudes are fully constrained. In the notation of the mixture model discussion above, we adopt some specific prior $p_\alpha(\chi_1, \chi_2|\sigma) = \delta(\chi_1 - x_\alpha)\delta(\chi_2 - y_\alpha)$ where $x_\alpha, y_\alpha$ are the spin $\lambda_\alpha$. A mixture model allowing generic $\lambda$ and thus including all such components allows both component spins to take arbitrary (discrete) values. [We could similarly extend our mixture model to include kicks.] The posterior distribution over all possible spin distributions $p(\lambda|d) = p(d|\lambda)p(\lambda)/p(d)$ follows from Bayes' theorem and the concrete likelihood given in Eq. (B.5). In practice, however, we don't generally compute or report the full posterior distribution, as it contains far more information than we need (e.g., the extent of the ensemble of possible spin distributions that fit the data). Instead, we compute the *expected spin distribution*

$$p_{post}(x) = \sum_\alpha \langle\lambda_\alpha\rangle p_\alpha \tag{B.6}$$

and the *variance in each* $\lambda_\alpha$. For the modest number of mixture components of interest here ($\simeq$ 100 possible choices of both spin magnitudes) and the modest degree of the polynomial ($\simeq$ 4−5), all necessary averages can be computed by direct symbolic quadrature of a polynomial in $\lambda_\alpha$. The integral can be expressed as a sum of terms of homogeneous degree in $\lambda$, and integrals of each of these terms can be carried out via the following general formula:

$$n! \int_{\sum_i x_i \le 1} dx_1 \dots dx_n x_{i_1}^{\alpha_1} \dots x_{i_Z}^{\alpha_Z} = \frac{n!}{(n-Z)!} \prod_{k=1}^{Z} B(\alpha_k + 1, n + 1 - k + \sum_{q>k} \alpha_q) \tag{B.7}$$

where the integral is over the region $x_i \ge 0$ and $\sum_i x_i \le 1$. We can also find the maximum likelihood estimate of $\lambda_\alpha$, for example by using the expectation-maximization algorithm [120]. In this work, however, we have many more basis models $\alpha = 1, 2, \dots$ used in our (spin) mixture than observations. Normally, we would reduce the effective dimension, for example by adopting prior assumptions in how the mixture coefficients can change as a function of spins $\chi_1, \chi_2$. To minimize additional formal overhead, we instead simply treat the spins hierarchically in blocks [Eq. (2.3)], considering lower-dimensional models where (for example) $\lambda_A$ denotes the a priori probability for $\chi_i \leq 0.6$ and $1-\lambda_A$ denotes the a priori probability for $\chi_i > 0.6$, so for example the prior probability for $(\chi_1, \chi_2) = 0.1$ is $\lambda_A^2/36$. In this four-block and one-parameter model, we can compute the average value of $\lambda_A$ in terms of the net weights associated with each block: $C_{AA,k} = \sum_{\chi_1,\chi_2\in A} c_{\alpha,k}$, $C_{A\bar{A}k}$, $C_{\bar{A}Ak}$ and $C_{\bar{A}\bar{A}k}$. For example, if for each of three synthetic observations, $C_{AA} = 1$ and all other weights are negligible, then we would conclude a posterori that $\langle\lambda_A\rangle = 0.875$ and $\sigma_{\lambda_A} = 0.11$. This approach was adopted in Figuure 2.7, in contrast to the preceding figures which adopted fixed natal spins for all BHs.

## B.3 Approximate posterior distribution for GW170104

For most events examinined in this study, we made use of posterior samples provided and performed by the LIGO Scientific Collaboration, generated by comparing each event to the IMRPv2 approximation [82]. Because we cannot employ the same level of detail for GW170104, we instead resort to an approximate posterior distribution, derived from the reported GW170104 results [3] and our understanding of gravitational wave parameter estimation, as approximated using a Fisher matrix [123].

For GW170104 we construct an approximate (truncated) Gaussian posterior distribution in only three correlated binary parameters: $\mathcal{M}_c, \eta, \chi_{\rm eff}$. The shape of this Gaussian (i.e., its inverse covariance matrix) was constructed via a Fisher matrix approximation, derived using the median detector-frame parameters reported for GW170104 (i.e., $m_1 \simeq 37.1 M_\odot$, $m_2 \simeq 22.6 M_\odot$, and – breaking dengeracy with an ad-hoc choice – $\chi_{1,z}\chi_{2,z} \simeq \chi_{\rm eff} \simeq -0.12$); the reported network SNR of GW170104 (i.e., $\rho \simeq 13.0$); and a suitable single-detector noise power spectrum. Our effective Fisher matrix estimate for the inverse covariance matrix $\Gamma$ [124] adopted the noise power spectrum at GW150914, using a minimum frequency $f_{\rm min} =$ 30Hz; employed the (nonprecessing) SEOBNRv4 approximation [125], evaluated on a grid in $\mathcal{M}_c, \eta, \chi_{1,z}, \chi_{2,z}$; and fit as a quadratic function of $\mathcal{M}_c, \eta, \chi_{\rm eff}$. We adopt a nonprecessing model and lower-dimensional Fisher matrix approximation because the posterior of this event, like GW150914, is consistent with nonprecessing spins and is very well approximated, in these parameters, by a nonprecessing model; see, e.g., [56]. This simple approximation captures important correlations between $\mathcal{M}_c, \eta$ and $\chi_{\rm eff}$, and the diagonal terms of $\Gamma^{-1}\rho^2$ roughly reproduce the width of the posterior distribution reported for GW170104. To obtain better agreement with the reported one-dimensional credible intervals, we scaled the terms $\Gamma_{\mathcal{M}_c,x}$ for $x = \mathcal{M}_c, \eta, \chi_{\rm eff}$ by a common scale factor 0.29 and the term $\Gamma_{\chi_{\rm eff},\chi_{\rm eff}}$ by 0.9. For similar reasons, we likewise hand-tuned the center of the Gaussian distribution to the (unphysical) parameter location to $\mathcal{M}_c = 22.9$, $\eta = 0.32$, $\chi_{\rm eff} = 0.013$. Using this ansatz, we generate GW170104-like posterior samples in $\mathcal{M}_c, \eta, \chi_{\rm eff}$ from this Gaussian distribution, truncating any unphysical samples (i.e., with $\eta > 1/4$). For our tuned posterior, the median and 90% credible regions on the synthetic posteriors approximate the values and ranges reported. According to our highly simplified and purely synthetic approach, the resulting 90% credible regions are $M_{\rm tot} = 51.2^{+7.6}_{-6.8} M_\odot$, $q = 0.62^{+0.25}_{-0.24}$, $\chi_{\rm eff} = -0.12^{+0.28}_{-0.27}$.